\documentclass[a4paper]{IEEEtran}

\usepackage[utf8]{inputenc}

\usepackage{amsmath,amssymb,cite,dsfont}
\usepackage{mathtools}
\usepackage{stmaryrd}	
\usepackage{bbm}	

\usepackage{enumerate}	

\usepackage[usenames,dvipsnames]{xcolor} 	
\usepackage{tikz}				
\usetikzlibrary{decorations.pathreplacing}	

\usepackage{pgfplots}
\pgfplotsset{compat=1.13}

\ifCLASSOPTIONcompsoc
\usepackage[caption=false,font=normalsize,labelfont=sf,textfont=sf]{subfig}
\else
\usepackage[caption=false,font=footnotesize]{subfig}
\fi

\usepackage{hyperref}
\hypersetup{
    colorlinks,%
    citecolor=black,%
    filecolor=black,%
    linkcolor=black,%
    urlcolor=black,%
    pdfauthor={German Bassi, Pablo Piantanida, and Shlomo Shamai (Shitz)},%
    pdftitle={The Wiretap Channel with Generalized Feedback}%
}
\usepackage[all]{hypcap}

\newcommand{\V}[1]{\ensuremath{\mathbf{#1}}}
\newcommand{\ve}[1]{\ensuremath{\underline{#1}}}

\newcommand{\PR}[1]{\ensuremath{\textnormal{Pr}\!\left\{{#1}\right\}}}

\newcommand{\I}[2]{\ensuremath{I(#1;#2)}}
\newcommand{\IC}[3]{\ensuremath{I(#1;#2\vert #3)}}
\renewcommand{\H}[1]{\ensuremath{H(#1)}}
\newcommand{\HC}[2]{\ensuremath{H(#1\vert #2)}}

\newcommand{\mkv}{-\!\!\!\!\minuso\!\!\!\!-}
\newcommand{\cond}{\,\vert\,}

\newcommand{\cC}{\ensuremath{\mathcal C}}

\newcommand{\cK}{\ensuremath{\mathcal K}}
\newcommand{\cM}{\ensuremath{\mathcal M}}

\newcommand{\cP}{\ensuremath{\mathcal P}}
\newcommand{\cQ}{\ensuremath{\mathcal Q}}
\newcommand{\cR}{\ensuremath{\mathcal R}}
\newcommand{\cS}{\ensuremath{\mathcal S}}
\newcommand{\cT}{\ensuremath{\mathcal T}}
\newcommand{\cU}{\ensuremath{\mathcal U}}
\newcommand{\cV}{\ensuremath{\mathcal V}}
\newcommand{\cX}{\ensuremath{\mathcal X}}
\newcommand{\cY}{\ensuremath{\mathcal Y}}
\newcommand{\cZ}{\ensuremath{\mathcal Z}}

\newcommand{\bM}{\ensuremath{\mathbb M}}

\newcommand{\bE}{\ensuremath{\mathbb E}}

\newcommand{\typ}[1]{\cT_\delta^n(#1)}
\newcommand{\typc}[1]{\cT_{\delta'}^n(#1)}

\newcommand{\ind}[1]{\ensuremath{\mathds{1} {\left\{ #1 \right\}}}}

\newtheorem{definition}	{Definition}
\newtheorem{theorem}	{Theorem}
\newtheorem{proposition}{Proposition}
\newtheorem{lemma}	{Lemma}

\newtheorem{remark}	{Remark}

\title{The Wiretap Channel with Generalized Feedback: Secure Communication and Key Generation}
\author{
  \IEEEauthorblockN{Germ{\'a}n~Bassi,~\IEEEmembership{Member,~IEEE}, Pablo~Piantanida,~\IEEEmembership{Senior Member,~IEEE}, and \\ Shlomo~Shamai~(Shitz),~\IEEEmembership{Fellow Member,~IEEE}}\\
  \thanks{This work was partially supported by the FP7 Network of Excellence in Wireless COMmunications NEWCOM\#.
  The work of G.~Bassi was funded in part by the Knut and Alice Wallenberg foundation and the Swedish Foundation for Strategic Research,  
  and the work of S.~Shamai was also supported by the European Union's Horizon 2020 Research And Innovation Programme, grant agreement no. 694630.
  The material in this paper was presented in part at the 
  2015 IEEE Information Theory Workshop (ITW), Oct. 2015~\cite{bassi_itw_2015}.}
  \thanks{G.~Bassi was with the Laboratoire des Signaux et Syst{\`e}mes (L2S, UMR CNRS 8506) CentraleSup{\'e}lec--CNRS--Universit{\'e} Paris-Sud, F-91192 Gif-sur-Yvette, France. He is now with the School of Electrical Engineering and Computer Science, KTH Royal Institute of Technology, Stockholm 100 44, Sweden (e-mail: germanb@kth.se).}
  \thanks{P.~Piantanida is with CentraleSup{\'e}lec--French National Center for Scientific Research (CNRS)--Universit{\'e} Paris-Sud, 3 Rue Joliot-Curie, F-91192 Gif-sur-Yvette, France, and with Montreal Institute for Learning Algorithms (MILA) at Universit{\'e} de Montr{\'e}al, 2920 Chemin de la Tour, Montr{\'e}al, QC H3T 1N8, Canada (e-mail: pablo.piantanida@centralesupelec.fr).}
  \thanks{S.~Shamai~(Shitz) is with the Department of Electrical Engineering, Technion--Israel Institute of Technology, Haifa, 32000, Israel (e-mail: sshlomo @ee.technion.ac.il).}
  \thanks{Copyright (c) 2018 IEEE. Personal use of this material is permitted. However, permission to use this material for any other purposes must be obtained from the IEEE by sending a request to pubs-permissions@ieee.org.}
}

\begin{document}

\maketitle

\begin{abstract}
It is a well-known fact that feedback does not increase the capacity of point-to-point memoryless channels, however, its effect in secure communications is not fully understood yet. In this work, an achievable scheme for the wiretap channel with generalized feedback is presented.
This scheme, which uses the feedback signal to generate a shared secret key between the legitimate users, encrypts the message to be sent at the bit level. New capacity results for a class of channels are provided, as well as some new insights into the secret key agreement problem.
Moreover, this scheme recovers previously reported rate regions from the literature, and thus it can be seen as a generalization that unifies several results in the field.
\end{abstract}

\begin{IEEEkeywords}
Information-theoretic security, wiretap channel, feedback, secret key, secrecy capacity, secret key capacity.
\end{IEEEkeywords}

\section{Introduction}

\IEEEPARstart{I}{n recent} years, there has been great interest in the study of 
the wiretap channel~(WTC)~\cite{liang_its_2008} as a model for secure 
communications against eavesdroppers by harnessing the randomness inherently 
present in the physical medium (see~\cite{bassily_cooperative_2013} and 
references therein).
Application to secure wireless networks is extremely attractive, not only 
because the open nature of the medium makes communication devices particularly 
sensitive to eavesdropping, but also because randomness is abundantly available 
in such scenarios.
As a matter of fact, the current theory of physical layer security indicates 
that the part of the data that is secured cannot be retrieved by the 
eavesdropper, regardless of its computational power.

A crucial observation behind this promising result is that unless the 
legitimate's and the eavesdropper's channels enjoy different statistical 
properties, which is often a nonrealistic assumption, secrecy cannot be 
guaranteed.
Nevertheless, if both channels share the same statistical properties but some 
extra outdated side information is available at the transmitter, then the 
encoder can create the asymmetry required to ensure security (e.g., 
see~\cite{yang_secrecy_2013, tandon_miso_2014}).
In fact, this observation reveals one of the major limitations of the wiretap 
model, whose performance strongly depends on the amount of outdated side 
information that may be available at the transmitter.
Studying the impact on secrecy systems of different types of instantaneous 
information is therefore of both practical and theoretical interest.

\begin{figure}
\centering
\begin{tikzpicture}[line width=1pt, font=\footnotesize]
\draw[-latex,thick] (0.4,2.6) -- (1,2.6) node[midway,above] {$\bM_n$};
\draw[rounded corners=4pt] (1,2.2) rectangle (2,3) node[pos=.5] {Alice};
\draw[-latex,thick] (2,2.6) -- (2.7,2.6) node[midway,above] {$X_i$};
\draw (2.7,2.2) rectangle (4.2,3) node[pos=.5] {$p(y \hat{y} z\vert x)$};
\draw[-latex,thick] (4.2,2.6) -- (4.9,2.6) node[midway,above] {$Y_i$};
\draw[-latex,thick] (3.45,2.2) -- (3.45,1.5) -- (4.9,1.5) node[midway,above] {$Z_i$};
\draw[-latex,thick] (3.45,3) -- (3.45,3.6) -- (1.5,3.6) -- (1.5,3) node[midway,xshift=-12pt,yshift=2pt] {$\hat{Y}^{i-1}$};
\draw[rounded corners=4pt] (4.9,2.2) rectangle (5.8,3) node[pos=.5] {Bob};
\draw[-latex,thick] (5.8,2.6) -- (6.4,2.6) node[midway,above] {$\hat{\bM}_n$} -- (6.45,2.6) node[right,xshift=-2pt] {$\PR{\hat{\bM}_n \!\neq\! \bM_n\!} \leq \epsilon$};
\draw[rounded corners=4pt] (4.9,1.1) rectangle (5.8,1.9) node[pos=.5] {Eve};
\draw[-latex,thick] (5.8,1.5) -- (6.45,1.5) node[right,xshift=-2pt] {$\,\IC{\bM_n}{Z^n}{\mathsf{c}_n} \leq \epsilon$};
\end{tikzpicture}
\caption{Wiretap channel with generalized feedback.
\label{fig:model}
}
\end{figure}
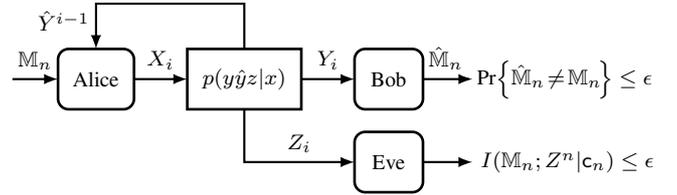

In this work, we investigate the problem where a node, Alice, wishes to secretly 
communicate a message to another node, Bob, in presence of a passive 
eavesdropper, Eve, as depicted in Fig.~\ref{fig:model}.
Alice can communicate with Bob using a general memoryless channel but Eve is 
listening this communication through another memoryless channel, whose 
statistical properties can be different or equal to Bob's. 
In addition, we assume that Alice observes general --may be noisy-- outdated 
feedback which is correlated to the channel outputs of Bob and Eve, referred to 
as ``generalized feedback''. 
It is worth mentioning that this feedback model is rich enough since it handles 
several different types of outdated side information at the transmitter (e.g., 
delayed state-feedback or noisy feedback of the channel outputs) as well as both 
\emph{secure} and \emph{non-secure feedback} scenarios.
Therefore, the generalized feedback model provides the adequate framework to 
investigate the impact of the feedback model.

\subsection{Related Work}

There has been substantial work on the wiretap channel with different feedback
models, however, the capacity in the general case remains unresolved.
Feedback, even partial, is known to increase the capacity of several 
multi-terminal networks with respect to the non-feedback case (e.g., 
broadcast~\cite{dueck_partial_1980} and multiple access 
channels~\cite{cover_achiev_1981}).
The transmitter uses the feedback signal to provide the decoder with noisy 
functions of the channel noise or parameters, and the messages.
This communication is accomplished by two fundamentally different classes of 
coding schemes: those based on block Markov (digital) 
coding~\cite{dueck_partial_1980, cover_achiev_1981},
and those based on linear (analog) encoding~\cite{schalkwijk_coding_1966}, known 
as \emph{Schalkwijk-Kailath} (S-K) scheme, which perform well over additive
Gaussian models.

In the literature, there exist two complementary approaches on the use of the
feedback signal to secure the communication.
On the first one, Alice and Bob extract common randomness from their respective
channel outputs which they use as a shared \emph{secret key}. This key encrypts
the message at the bit level which provides secrecy as long as Eve cannot obtain
the key.
On the second approach, Alice relies on a ``feedback-dependent codebook'' that
correlates the codewords to be sent with the feedback signal. In this way, Alice
seeks to hide as much as possible the transmitted codewords from Eve's
observations (e.g., \emph{beamforming} at the codeword level).
Due to the inherently digital nature of encrypting the message bitwise, only the
block Markov scheme is suited for the first approach, while both block Markov
and S-K schemes are possible for the second methodology.

Results based on the secret key approach are numerous, as it seems natural to
use the feedback link (secure or not) to agree upon a key.
In~\cite{ahlswede_trans_2006}, the authors analyze the WTC with perfect
output feedback only at the encoder and propose a scheme based on this 
methodology. This scheme achieves the capacity of the \emph{degraded}, i.e., 
$X\mkv Y\mkv Z$, and \emph{reversely degraded}, i.e., $X\mkv Z\mkv Y$, WTC with 
perfect output feedback. 
The case of parallel channels, i.e., $Y\mkv X\mkv Z$, is studied
in~\cite{dai_capacity_2012}, where the secrecy capacity is characterized when 
one of the channels is \emph{more capable} than the other.
A similar model to~\cite{ahlswede_trans_2006}, where the feedback link is in
fact a secure rate-limited channel from Bob to Alice, is presented
in~\cite{ardestanizadeh_wiretap_2009}. In contrast to the previous schemes, the
key is here created with \emph{fresh} randomness that Bob transmits.

The use of state-feedback as a means to generate a key has also been analyzed, 
either when it is known only by the legitimate users~\cite{chia_wiretap_2012} or 
by all the nodes in the network~\cite{czap_secret_2015}. The authors 
of~\cite{chia_wiretap_2012} propose a lower bound for the general discrete 
memoryless WTC with state information at both the encoder and decoder, 
which is tight in several scenarios, e.g., when Bob is \emph{less noisy} than 
Eve, or when Eve is less noisy than Bob and the channel is independent of the 
state. 
In~\cite{czap_secret_2015}, the authors study a 
communication scenario where an encoder transmits private messages to several 
receivers through a broadcast erasure channel, and the receivers feed back 
(publicly) their channel states. Capacity is characterized based on linear 
complexity two-phase schemes: in the first phase appropriate secret keys are 
generated which are exploited during the second phase to encrypt each message.

Indeed, the generation of the secret key is a problem in and of 
itself~\cite{ahlswede_common_1993, maurer_secret_1993}. 
Two models exist that tackle this issue: the ``source model'', when the 
generation is based on the common randomness present in correlated sources, and 
the ``channel model'', when the common randomness is due to the correlation 
between inputs and outputs of a channel.
The authors of~\cite{csiszar_common_2000} study the first model, where two nodes 
generate common randomness with the aid of a third ``helper'' node, all of them 
connected by noiseless rate-limited links. This common randomness may be kept 
secret from a fourth passive node that acts as an eavesdropper.
The same authors also analyze the channel model in~\cite{csiszar_secrecy_2008}.
Capacity results are presented in both~\cite{csiszar_common_2000} 
and~\cite{csiszar_secrecy_2008} when there is only one round of 
communication over the noiseless public link.
General lower and upper bounds for both source and channel models when interaction
is allowed are found in~\cite{gohari_sk-source_2010, gohari_sk-channel_2010}.

More recently, \cite{khisti_secret-key_2012} investigates a similar problem 
as~\cite{csiszar_common_2000} but 
there is no helper node, the users communicate over a WTC, and a public 
discussion channel may or may not be available.
On the other hand, \cite{salimi_key_2013} analyzes key agreement over a multiple 
access channel, i.e., the channel model. Here the receiver can actively send 
feedback, through a noiseless or noisy link, to increase the size of the shared 
key.
The authors of~\cite{prabhakaran_secrecy_2012} go one step further and study 
the simultaneous transmission of a secret message along with a key 
generation scheme using correlated sources. They obtain a simple expression that 
shows the trade-off between the achievable secrecy rate and the achievable 
secret key rate.

Results based on the ``feedback-dependent codebook'' approach, however, are not 
that numerous to the best of our knowledge.
Early work in~\cite{tang_multiple_2007, ekrem_effects_2008} study the multiple 
access channel (MAC) with generalized feedback and secrecy constraints. 
In~\cite{tang_multiple_2007} the eavesdropper is an external user to the MAC and 
the cooperating encoders use (partial) \emph{decode-and-forward} strategies to 
enlarge their achievable rates. On the other hand, in~\cite{ekrem_effects_2008}, 
each encoder acts as an eavesdropper for the other user and the authors propose 
lower bounds based on \emph{compress-and-forward} to increase the transmission 
rates to levels that are only decodable by the destination. Completely outdated 
state-feedback can also be used to enhance security.  
In~\cite{yang_secrecy_2013, tandon_miso_2014}, it is shown that outdated 
state-feedback of either the legitimate channel, the eavesdropper's channel or 
both, increases the secure degrees of freedom of the two-user Gaussian 
multiple-input multiple-output (MIMO) wiretap channel.

Active feedback in a half-duplex fashion is used in~\cite{tung_secure_2010}, 
where communication is split in two phases. In the first one, the destination 
sends a random codeword which cannot be decoded by the eavesdropper. On top of 
this ``interference sequence'', the codeword to be transmitted in the second 
phase is superimposed. This scheme achieves positive secrecy rates in the MIMO 
wiretap channel even when the eavesdropper has  more antennas than the source. 
An analogous scheme is presented for the full-duplex two-way Gaussian wiretap 
channel in~\cite{he_role_2013}. Here, the interference sequence sent in the 
first phase is canceled at the eavesdropper thanks to the full-duplex operation 
of the channel. Moreover, the authors show that neglecting the feedback signal 
can lead to unbounded loss in achievable rate under certain conditions.

In~\cite{lai_wiretap_2008}, the modulo-additive WTC with a full-duplex 
destination node is investigated. The authors propose a scheme where the 
legitimate receiver injects noise in the backward (feedback) channel, 
effectively eliminating any correlation between the message sent and the 
eavesdropper's observation. This scheme achieves the full capacity of the 
point-to-point channel in absence of the wiretapper, i.e., full secrecy can be 
guaranteed at no rate cost. A similar conclusion is also drawn 
in~\cite{gunduz_secret_2008}, where the authors analyze an additive white 
Gaussian noisy (AWGN) channel with perfect output feedback from the legitimate 
receiver. They propose a S-K coding scheme which achieves the full 
capacity of the AWGN channel in absence of the wiretapper, as long as the 
eavesdropper has only access to a noisy feedback signal. This last result is 
generalized by the authors in~\cite{bassi_isit_2015}, where an achievable 
strategy that combines block Markov and S-K schemes is introduced.

A closely related topic to the one addressed in this work is the WTC with 
\emph{noncausal} side-information available to the parties. The model where the 
side-information is only available at the encoder is studied 
in~\cite{chen_wtc_2008}, where a lower bound based on Gelfand and Pinsker's 
strategy for channels with state~\cite{gelfand_coding_1980} is introduced. An 
extension to this model, with both the encoder and legitimate decoder having 
access to correlated side-information, is investigated 
in~\cite{liu_wiretap_2007}. More recently, the authors of~\cite{bafghi_itw_2012} 
analyze a slightly different scenario where the state affecting the legitimate 
decoder's channel is not equal to the one affecting the eavesdropper's channel. 
These channel states are correlated and the encoder only knows the state of the 
legitimate decoder's channel.

\subsection{Contributions and Organization of the Paper}

In this work, we derive the following results: 
\begin{itemize}
\item We first introduce our main contribution (see Theorem~\ref{theo-KG}), a lower bound based on the secret key approach, where the feedback link is used to generate a key that encrypts the message partially or completely.

\item As an extension of Theorem~\ref{theo-KG}, we derive a lower bound (see Theorem~\ref{cor-key}) on secret key agreement for the same channel model. The channel is used both as a source of correlated randomness and as a means of communication, i.e., there is no parallel public noiseless channel used by the terminals.

\item In order to assess the optimality of these strategies, we derive upper bounds for a particular class of channels (see Theorems~\ref{theo-OB} and~\ref{theo-KA-OB}) and we show that the lower bound and its extension are optimal under some special conditions (see Propositions~\ref{prop:wsk} to~\ref{prop:6}). 

\item In addition to these new capacity results, the first lower bound is shown to recover previously reported results for different channel and feedback models (see Theorems~\ref{theorem-5} and~\ref{theorem-6}). Consequently, the lower bound provided in this work can be seen as a generalization and thus unification of several results in the field.
\end{itemize}

The rest of this paper is organized as follows. Section~\ref{sec:Problem} introduces the general channel model and the one used for the capacity results, as well as some basic definitions. In Section~\ref{sec:Main}, we present our main results: the lower and upper bounds, whose proofs are deferred to the appendices. The new capacity results and the comparison with previously reported lower bounds are shown in Section~\ref{sec:Capacity}, while the summary and concluding remarks are stated in Section~\ref{sec:Summary}.

\subsection*{Notation and Conventions}

In this work, we use the standard notation of~\cite{gamal_network_2011}.
Specifically, given two integers $i$ and $j$, the expression $[i: j]$ denotes
the set $\{i, i+1, \ldots, j\}$, whereas for real values $a$ and $b$, $[a, b]$
denotes the closed interval between $a$ and $b$.
Lowercase letters such as $x$ and $y$ are mainly used to represent constants or
realizations of random variables, capital letters such as $X$ and $Y$ stand for
the random variables in itself, while calligraphic letters such as $\cX$ and 
$\cY$ are reserved for sets, codebooks or special functions.

We use the notation $x_i^j = (x_i, x_{i+1}, \ldots, x_j)$ to denote the
sequence of length $j-i+1$ for $1\leq i\leq j$. If $i=1$, we drop the subscript
for succinctness, i.e., $x^j = (x_1, x_2, \ldots, x_j)$.
For simplicity, $n$-sequences may be denoted either by $x^n$ or $\V{x}$. This 
comes in handy in the proofs where we deal with $b$ blocks of $n$-sequences, 
i.e., $\V{x}^b = (\V{x}_1, \V{x}_2, \ldots, \V{x}_b)$.

The probability distribution~(PD) of the random vector $X^n$, $p_{X^n}(x^n)$, is
succinctly written as $p(x^n)$ without subscript when it can be understood from
the argument $x^n$.
Given three random variables $X$, $Y$, and $Z$, if its joint PD can be
decomposed as $p(xyz) = p(x) p(y\vert x) p(z\vert y)$, then they form a Markov
chain, denoted by $X \mkv Y \mkv Z$.
Entropy is denoted by $\H{\cdot}$ and mutual information, $\I{\cdot}{\cdot}$. 
The expression $|x|^+$ stands for $\max\{ x, 0 \}$.
%

\section{Problem Definition}
\label{sec:Problem}

In this work, we consider primarily the wiretap channel with generalized feedback (WTC-GF). Nonetheless, we also provide some insights on a specific class of channels that can be derived from the original system model. We now introduce these two models.

\subsection{Wiretap Channel with Generalized Feedback}

In the WTC-GF, Alice wants to securely transmit a message $\bM_n$ (uniformly distributed over a message set $\cM_n$) to Bob with the aid of a feedback signal while Eve observes the transmission. The WTC-GF, depicted in Fig.~\ref{fig:model}, is modeled as a discrete memoryless channel whose $n$th extension satisfies
\begin{equation}
p(y_i \hat{y}_i z_i \vert x^i y^{i-1} \hat{y}^{i-1} z^{i-1}) = p(y_i \hat{y}_i z_i \vert x_i ) ,
\label{eq:wtc_nth_extension}
\end{equation}
for all $i\in[1:n]$. The right-hand side of~\eqref{eq:wtc_nth_extension} is independent of the time slot $i$ and it is defined by the conditional probability distribution
\begin{equation}
p(y \hat{y} z \vert x ):\, \cX \to \cY\times\hat{\cY}\times\cZ , 
\label{eq:wtc_model}
\end{equation}
where $x\in\cX$ is Alice's channel input, $\hat{y}\in\hat{\cY}$ is the feedback signal, and $y\in\cY$ and $z\in\cZ$ are Bob's and Eve's channel outputs, respectively.


\vspace{1mm}
\begin{definition}[Code]\label{def:s_code}
A $(2^{nR},n)$ code $\mathsf{c}_n$ for the WTC-GF consists of
a message set $\cM_n\triangleq[1:2^{nR}]$,
a source of local randomness at the encoder $R_r\in\cR_r$,
a family of encoding functions $\textrm{enc}_i: (\cM_n, \cR_r, \hat{\cY}^{i-1}) \to \cX_i$, and
a decoding function $\textrm{dec} : \cY^n \to \cM_n$.
\end{definition}
\vspace{1mm}

The reliability performance of the $(2^{nR},n)$ code $\mathsf{c}_n$ is measured in terms of its average probability of error
\begin{equation}
\mathsf{P}_{\!e}(\mathsf{c}_n) \triangleq \PR{ \textrm{dec}(Y^n) \neq \bM_n \vert \mathsf{c}_n} ,
\label{eq:def_rel}
\end{equation}
while its secrecy performance is measured in terms of the information leakage
\begin{equation}
\mathsf{L}(\mathsf{c}_n) \triangleq \IC{\bM_n}{Z^n}{\mathsf{c}_n} .
\label{eq:def_leak}
\end{equation}

\vspace{1mm}
\begin{definition}[Achievable Rate]\label{def:s_rate}
A \emph{weak secrecy rate} $R$ is achievable for the WTC-GF if for every $\epsilon>0$ and sufficiently large $n$, there exists a $(2^{nR},n)$ code $\mathsf{c}_n$ such that
\begin{align}
 \label{eq:s_rate_weak}
 \mathsf{P}_{\!e}(\mathsf{c}_n) &\leq \epsilon &
 &\textnormal{and} &
 \frac{1}{n}\mathsf{L}(\mathsf{c}_n) &\leq \epsilon.
\end{align}

On the other hand, a \emph{strong secrecy rate} $R$ is achievable for the WTC-GF if for every $\epsilon>0$ and sufficiently large $n$, there exists a $(2^{nR},n)$ code $\mathsf{c}_n$ such that
\begin{align}
 \label{eq:s_rate_strong}
 \mathsf{P}_{\!e}(\mathsf{c}_n) &\leq \epsilon &
 &\textnormal{and} &
 \mathsf{L}(\mathsf{c}_n) &\leq \epsilon.
\end{align}
\end{definition}
\vspace{1mm}

\begin{definition}[Capacity]\label{def:s_cap}
The \emph{weak secrecy capacity} $\textnormal{C}_{sf}$ of the WTC-GF is the supremum of all achievable weak secrecy rates.
Similarly, the \emph{strong secrecy capacity} $\overline{\textnormal{C}}_{sf}$ of the WTC-GF is the supremum of all achievable strong secrecy rates.
\end{definition}
\vspace{1mm}

In this work, we also consider the situation where the source does not want to
transmit a message but rather agree on a \emph{secret key} (SK) with the legitimate decoder
while keeping it private from the eavesdropper. The channel outputs, i.e., $y$,
$\hat{y}$, and $z$, may be seen as correlated sources. This scenario is called
``channel model'' for key agreement, but in our case, the communication also
takes place in the same channel rather than in a separate noiseless public
broadcast channel.

\vspace{1mm}
\begin{definition}[SK Code]\label{def:sk_code}
A $(2^{nR_k},n)$ secret key code $\mathsf{c}_n$ for the WTC-GF consists of
a key set $\cK_n\triangleq[1:2^{nR_k}]$,
a source of local randomness at the encoder $R_r\in\cR_r$,
a family of encoding functions $\varphi_i: (\cR_r, \hat{\cY}^{i-1}) \to \cX_i$,
a key generation function $\psi_a: (\cR_r, \hat{\cY}^n) \to \cK_n$, and
a key generation function $\psi_b: \cY^n \to \cK_n$.
\end{definition}
\vspace{1mm}

Let $K=\psi_a(R_r, \hat{Y}^n)$, then, similar to~\eqref{eq:def_rel}--\eqref{eq:def_leak}, the performance of the $(2^{nR_k},n)$ secret key code $\mathsf{c}_n$ is measured in terms of its average probability of error
\begin{equation}
\mathsf{P}_{\!e}(\mathsf{c}_n) \triangleq \PR{ \psi_b(Y^n) \neq K \vert \mathsf{c}_n},
\label{eq:def_sk_rel}
\end{equation}
in terms of the information leakage
\begin{equation}
\mathsf{L}_k(\mathsf{c}_n) \triangleq \IC{K}{Z^n}{\mathsf{c}_n} ,
\label{eq:def_sk_leak}
\end{equation}
and in terms of the uniformity of the keys
\begin{equation}
\mathsf{U}_k(\mathsf{c}_n) \triangleq nR_k -\HC{K}{\mathsf{c}_n} .
\label{eq:def_sk_unif}
\end{equation}

\vspace{1mm}
\begin{definition}[Achievable SK Rate]\label{def:sk_rate}
A \emph{weak secret key rate} $R_k$ is achievable for the WTC-GF if for every $\epsilon>0$ and sufficiently large $n$, there exists a $(2^{nR_k},n)$ SK code $\mathsf{c}_n$ such that
\begin{align}
 \label{eq:sk_rate_weak}
 \mathsf{P}_{\!e}(\mathsf{c}_n) &\leq \epsilon , &
 \frac{1}{n}\mathsf{L}_k(\mathsf{c}_n) &\leq \epsilon , &
 & \textnormal{and} &
 \frac{1}{n}\mathsf{U}_k(\mathsf{c}_n) &\leq \epsilon .
\end{align}

On the other hand, a \emph{strong secret key rate} $R_k$ is achievable for the WTC-GF if for every $\epsilon>0$ and sufficiently large $n$, there exists a $(2^{nR_k},n)$ SK code $\mathsf{c}_n$ such that
\begin{align}
 \label{eq:sk_rate_strong}
 \mathsf{P}_{\!e}(\mathsf{c}_n) &\leq \epsilon , &
 \mathsf{L}_k(\mathsf{c}_n) &\leq \epsilon , &
 & \textnormal{and} &
 \mathsf{U}_k(\mathsf{c}_n) &\leq \epsilon .
\end{align}
\end{definition}
\vspace{1mm}

\begin{definition}[SK Capacity]\label{def:sk_cap}
The \emph{weak secret key capacity} $\textnormal{C}_{kf}$ of the WTC-GF is the supremum of all achievable weak SK rates.
Similarly, the \emph{strong secret key capacity} $\overline{\textnormal{C}}_{kf}$ of the WTC-GF is the supremum of all achievable strong SK rates.
\end{definition}
\vspace{1mm}

\subsection{Wiretap Channel with Parallel Sources}

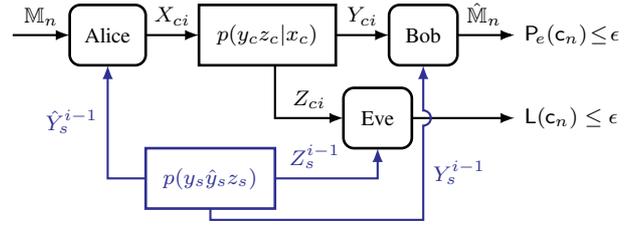
\begin{figure}[t!]
\centering
\begin{tikzpicture}[line width=1pt, font=\footnotesize]
\draw[-latex,thick] (0.25,2.6) -- (1,2.6) node[midway,above] {$\bM_n$};
\draw[rounded corners=4pt] (1,2.2) rectangle (2,3) node[pos=.5] {Alice};
\draw[-latex,thick] (2,2.6) -- (2.7,2.6) node[midway,above] {$X_{ci}$};
\draw (2.7,2.2) rectangle (4.5,3) node[pos=.5] {$p(y_c z_c\vert x_c)$};
\draw[-latex,thick] (4.5,2.6) -- (5.2,2.6) node[midway,above] {$Y_{ci}$};
\draw[-latex,thick] (3.7,2.2) -- (3.7,1.5) -- (4.6,1.5) node[midway,above] {$Z_{ci}$};
\draw[rounded corners=4pt] (5.2,2.2) rectangle (6.1,3) node[pos=.5] {Bob};
\draw[-latex,thick] (6.1,2.6) -- (6.8,2.6) node[midway,above] {$\hat{\bM}_n$} -- (6.85,2.6) node[right] {$\mathsf{P}_{\!e}(\mathsf{c}_n) \!\leq\! \epsilon$};
\draw[rounded corners=4pt] (4.6,1.1) rectangle (5.5,1.9) node[pos=.5] {Eve};
\draw[-latex,thick] (5.5,1.5) -- (6.85,1.5) node[right] {$\mathsf{L}(\mathsf{c}_n) \leq \epsilon$};
\draw[Blue] (2,0.3) rectangle (3.7,1.1) node[pos=.5] {$p(y_s \hat{y}_s z_s)$};
\draw[-latex,thick,Blue] (2,.7) -- (1.5,.7) -- (1.5,2.2) node[left,midway] {$\hat{Y}_s^{i-1}$};
\draw[-latex,thick,Blue] (3.7,.7) -- (4.75,.7) node[above,midway] {$Z_s^{i-1}$} -- (5.05,.7) -- (5.05,1.1);
\draw[-latex,thick,Blue] (2.85,.3) -- (2.85,.15) -- (5.65,.15) -- (5.65,1.4) node[right,midway] {$Y_s^{i-1}$} arc (-90:90:.1) -- (5.65,2.2);
\end{tikzpicture}
\caption{Wiretap channel with independent correlated sources.
\label{fig:model_parallel}}
\end{figure}

The channel model~\eqref{eq:wtc_model} is general enough to encompass different special scenarios; one of them, that we use later in the derivation of our capacity results, is depicted in Fig.~\ref{fig:model_parallel}. This model is a WTC without channel feedback where each node has causal access to correlated sources; in particular, Alice, Bob, and Eve observe $\hat{Y}_s$, $Y_s$, and $Z_s$, respectively. The sources are i.i.d. and independent of the main channel's variables $(X_c,Y_c, Z_c)$. The new model may thus be defined based on the original one by the specific set of variables
\begin{subequations}\label{eq:parallel_var}
\begin{align}
 \hat{Y} &\triangleq \hat{Y}_s , &
 Y &\triangleq (Y_s, Y_c) , &
 & \textnormal{and} &
 Z &\triangleq (Z_s, Z_c) ,
\end{align}
with the following probability distribution
\begin{equation}
p(y_s y_c \hat{y}_s z_s z_c \vert x_c) = p(y_c z_c \vert x_c) p(y_s \hat{y}_s z_s) . 
\end{equation}
\end{subequations}

The performance metrics~\eqref{eq:def_rel}--\eqref{eq:def_leak} and~\eqref{eq:def_sk_rel}--\eqref{eq:def_sk_unif} 
as well as Definitions~\ref{def:s_code}--\ref{def:sk_cap} for the problems of weak secrecy capacity ($\textnormal{C}_{s}$), strong secrecy capacity ($\overline{\textnormal{C}}_{s}$), weak secret key capacity ($\textnormal{C}_{k}$), and strong secret key capacity ($\overline{\textnormal{C}}_{k}$), may be readily extended to this new model using the set of variables~\eqref{eq:parallel_var}.

\section{Summary of Main Results}
\label{sec:Main}

We present the main results of this work in the sequel. The proofs of these results are deferred to the appendices.

\subsection{Wiretap Channel with Generalized Feedback}

\subsubsection{Secrecy Rate Lower Bound}

We first introduce our main contribution, a coding scheme that allows Alice and Bob to agree on a secret key simultaneously with the transmission of a message. The secret key is generated by virtue of the feedback link and is used to encrypt at the bit level the next message to be sent.
For ease of reference, the achievable scheme is denoted as ``KG lower bound''.

\vspace{1mm}
\begin{theorem}[KG Lower Bound]\label{theo-KG}
A lower bound on the \emph{strong secrecy capacity} of the WTC-GF is given by
\begin{equation*}
\overline{\textnormal{C}}_{sf} \geq \max\left\{ \,\max_{p\in\cP_{I_1}} R_{KG_1}(p), \ \max_{p'\in\cP_{I_2}} R_{KG_2}(p') \right\}, \label{eq:theo_KG-rate}
\end{equation*}
where $R_{KG_1}(p)$ is the set of all nonnegative rates satisfying%
\begin{subequations}\label{eq:KG_region1_rate}
\begin{align}
 R_{KG_1} &\leq \I{U}{Y} -\IC{U}{Z}{Q} -\IC{U}{T}{QZ}  \nonumber\\
&\quad -\max\{ \I{Q}{Y}, \, \IC{V}{X\smash{\hat{Y}}}{UY} \} \nonumber\\ 
&\quad +\IC{V}{Y}{UT}-\IC{V}{Z}{UT} , \label{eq:KG_region1_rate1}\\
 R_{KG_1} &\leq \I{U}{Y} -\max\{ \I{Q}{Y}, \, \IC{V}{X\hat{Y}}{UY} \} , \label{eq:KG_region1_rate2}
\end{align}
\end{subequations}
whereas $R_{KG_2}(p')$ is the set of all nonnegative rates satisfying%
\begin{subequations}\label{eq:KG_region2_rate}
\begin{align}
R_{KG_2} &\leq \IC{V}{Y}{UT} -\IC{V}{Z}{UT} , \label{eq:KG_region2_rate1} \\
R_{KG_2} &\leq \I{U}{Y} -\IC{V}{X\hat{Y}}{UY} . \label{eq:KG_region2_rate2}
\end{align}
\end{subequations}
The maximization is performed over $\cP_{I_1}$, the set of all probability distributions given by
\begin{multline}
 \cP_{I_1} = \big\{ \, p( quxvty\hat{y}z ) =\\ p(qu) p( x \vert u)  p( y \hat{y} z \vert
x ) p( t \vert v ) p( v \vert u x \hat{y} ) \,\big\} , \label{eq:KG_pmf}
\end{multline}
and $\cP_{I_2}$, the subset in $\cP_{I_1}$ with $Q=\emptyset$. In the maximization, it suffices to consider $|\cQ|\leq|\cX|+4$, $|\cU|\leq|\cQ|(|\cX|+3)$, $|\cT|\leq|\cX|\cdot|\hat{\cY}|+2$, and $|\cV|\leq|\cT|(|\cX|\!\cdot\!|\hat{\cY}|+1)$.
\end{theorem}
\begin{IEEEproof}
In this scheme, the transmission is split into several blocks and the
transmitted message in each block is encrypted fully ($R_{KG_2}$) or
partially ($R_{KG_1}$).
The codewords $\V{T}$ and $\V{V}$ are used to convey a description of the feedback signal $\V{\hat{Y}}$ from the previous block, and thus they allow the legitimate users to \emph{generate} the secret key during the transmission. In $R_{KG_1}$, the description is sent partially by $\V{Q}$ and $\V{U}$, hence the presence of the maximum in~\eqref{eq:KG_region1_rate}.
Refer to Appendix~\ref{sec:Proof_KG} for further details.
\end{IEEEproof}
\vspace{1mm}

Insights behind~\eqref{eq:KG_region1_rate} may be found by rewriting it as%
\begin{subequations}
\begin{align}
R_{KG_1} &\leq \IC{U}{Y}{Q} -\IC{U}{Z}{Q} -\IC{U}{T}{QZ} \nonumber\\
 &\quad +\IC{V}{Y}{UT} -\IC{V}{Z}{UT} , \label{eq:KG_region1bis_rate1} \\
 R_{KG_1} &\leq \IC{U}{Y}{Q} , \label{eq:KG_region1bis_rate2}
\end{align}
subject to
\begin{equation}
\IC{V}{X\hat{Y}}{UY}\leq \I{Q}{Y} . \label{eq:KG_region1bis_rate3}
\end{equation}
\end{subequations}
The achievable secrecy rate~\eqref{eq:KG_region1bis_rate1} has two main components: a part due to Wyner's wiretap coding scheme, given by the first two terms, and a part due to the encrypted message, given by the last two terms in~\eqref{eq:KG_region1bis_rate1}. The remaining term, i.e., $\IC{U}{T}{QZ}$, represents a rate penalty due to the correlation between the channel codeword $\V{U}$ and the description $\V{T}$ that Eve decodes.
Moreover, the achievable secrecy rate cannot be larger than the ``effective link capacity''~\eqref{eq:KG_region1bis_rate2}, i.e., the link capacity $\I{U}{Y}$ once the cost of the key agreement scheme~\eqref{eq:KG_region1bis_rate3} is subtracted.

A similar analysis may be performed with~\eqref{eq:KG_region2_rate}, where only an encrypted message is sent.

\vspace{1mm}
\begin{remark}
If we set $Q=T=V=\emptyset$, we recover the achievable secrecy rate of the WTC without feedback.
\end{remark}
\vspace{1mm}

\subsubsection{SK Rate Lower Bound}

In the absence of a message, the scheme in Theorem~\ref{theo-KG} may
be employed by Alice and Bob to agree upon a secret key. This key could later
be used to encrypt the transmission or part of it on a higher layer.

\vspace{1mm}
\begin{theorem}\label{cor-key}
A lower bound on the \emph{strong secret key capacity} of the WTC-GF is given by 
\begin{align}
\overline{\textnormal{C}}_{kf} \geq  \smash{\max\limits_{p\in\cP_{I_1}}} &\Big[ \IC{V}{Y}{UT} -\IC{V}{Z}{UT} +  \big| \I{U}{Y} \nonumber\\
&\ -\max\{ \I{Q}{Y}, \, \IC{V}{X\hat{Y}}{UY} \} \nonumber\\[1mm]
&\ -\IC{U}{Z}{Q} -\IC{U}{T}{QZ} \big|^+ \Big] ,
\label{eq:cor-key_rate}
\end{align}
subject to
\begin{equation}
\IC{V}{X\hat{Y}}{UY} \leq \I{U}{Y} . \label{eq:cor-key_cond}
\end{equation}
The maximization in~\eqref{eq:cor-key_rate} is performed over $\cP_{I_1}$, defined in~\eqref{eq:KG_pmf}, and it suffices to consider random variables with the same bounded cardinalities as in Theorem~\ref{theo-KG}.
\end{theorem}
\begin{IEEEproof}
This result is a special case of the strategy in Theorem~\ref{theo-KG}, where
there is no message to be transmitted, i.e., $R=0$, and we are only interested
in generating a secret key. Refer to Appendix~\ref{sec:Proof_Cor-KG} for details.
\end{IEEEproof}

\vspace{1mm}
\begin{remark}
The results of Theorems~\ref{theo-KG} and~\ref{cor-key} are obtained using the \emph{weak} secrecy conditions~\eqref{eq:s_rate_weak} and~\eqref{eq:sk_rate_weak}, respectively.
However, employing the method introduced in~\cite{maurer_strong_2000}, we can show that the \emph{strong} secrecy conditions~\eqref{eq:s_rate_strong} and~\eqref{eq:sk_rate_strong} also hold true; therefore the theorems are expressed in terms of these stronger notions of secrecy.
\end{remark}


\subsection{Wiretap Channel with Parallel Sources}

\subsubsection{Secrecy Rate Upper Bound for a Class of Channels}

For the specific channel model depicted in Fig.~\ref{fig:model_parallel}, we derive the following upper bound on the secrecy capacity.

\vspace{1mm}
\begin{theorem}\label{theo-OB}
An upper bound on the \emph{strong secrecy capacity} of the wiretap channel with parallel sources is given by
\begin{equation}
 \overline{\textnormal{C}}_{s} \leq \max\limits_{p\in\cP_{o}} R ,
\end{equation}
where $R$ is a nonnegative rate satisfying
\begin{subequations}\label{eq:theo_OB-rate}
\begin{flalign}
\ R &\leq \I{U}{Y_c} -\I{U}{Z_c} +\IC{V}{Y_s}{T} -\mathrlap{\IC{V}{Z_s}{T} ,} &\label{eq:theo_OB-rate1}\\
\ R &\leq \I{X_c}{Y_c} -\IC{V}{\hat{Y}_s}{Y_s} , &\label{eq:theo_OB-rate2}
\end{flalign}
\end{subequations}
and the set of all input probability distributions is given by%
\begin{multline}
\cP_{o} = \big\{ \, p( ux_c vt y_c z_c y_s \hat{y}_s z_s ) = \\
p( u x_c ) p(y_c z_c \vert x_c) p(y_s \hat{y}_s z_s) p( t \vert v ) p( v \vert \hat{y}_s ) \, \big\} ,
\label{eq:theo-OB-pmf}
\end{multline}
with $|\cU| \leq |\cX_c|$, $|\cT| \leq |\hat{\cY}_s| +1$, and $|\cV| \leq (|\hat{\cY}_s| +1)^2$.
\end{theorem}
\begin{IEEEproof}
Refer to Appendix~\ref{sec:Proof-OB}.
\end{IEEEproof}

\vspace{1mm}
\begin{remark}
In the absence of the correlated sources, the bound~\eqref{eq:theo_OB-rate} collapses to the upper bound of the wiretap channel.
\end{remark}
\vspace{1mm}

\subsubsection{SK Rate Upper Bound for a Class of Channels}

Let us now consider that, in the scenario depicted in Fig.~\ref{fig:model_parallel},
Alice and Bob want to agree upon a secret key by means of the correlated
sources and the communication through the wiretap channel.

\vspace{1mm}
\begin{theorem}\label{theo-KA-OB}
An upper bound on the \emph{strong secret key capacity} of this channel model is given by
\begin{equation}
\overline{\textnormal{C}}_{k} \leq \max_{p\in\cP_o} \big[ \I{U}{Y_c} -\I{U}{Z_c} +\IC{V}{Y_s}{T} -\IC{V}{Z_s}{T} \big] , \label{eq:theo_KA-OB-rate}
\end{equation}
subject to
\begin{equation}
\IC{V}{\hat{Y}_s}{Y_s} \leq \I{X_c}{Y_c} , \label{eq:theo_KA-OB-cond}
\end{equation}
where the set of all input probability distributions $\cP_o$ is defined in~\eqref{eq:theo-OB-pmf} and the auxiliary random variables have the same bounded cardinalities as in Theorem~\ref{theo-OB}.
\end{theorem}
\begin{IEEEproof}
Refer to Appendix~\ref{sec:Proof_KA-OB}.
\end{IEEEproof}

\vspace{1mm}
\begin{remark}
The upper bound~\eqref{eq:theo_KA-OB-rate} is the sum of the secrecy capacity of the wiretap
channel $p(y_c z_c \vert x_c)$, the first two terms on the right-hand side
of~\eqref{eq:theo_KA-OB-rate}, and the secret key capacity of the WTC with a
public noiseless channel and correlated sources~\cite[Thm.
2.6]{csiszar_common_2000}, the other two terms in~\eqref{eq:theo_KA-OB-rate}.
\end{remark}

\vspace{1mm}
\begin{remark}
Although the upper bounds in Theorems~\ref{theo-OB} and~\ref{theo-KA-OB} are derived under the assumption that Alice observes its source sequence causally, both upper bounds are valid even if Alice has \emph{noncausal} access to it.
\end{remark}

\section{Capacity Results for Some Channel and Feedback Models}
\label{sec:Capacity}

In this section, we first introduce new capacity results for the wiretap channel with parallel sources obtained by the KG lower bound (Sections~\ref{ssec:Cap_SK} and~\ref{ssec:Cap_Secrecy}).
Next, we show that previously reported results for other types of channel and feedback models are recovered by this scheme as well (Sections~\ref{ssec:Cap_WTC-POF} and~\ref{ssec:Cap_WTC-CSI}).
Finally, we present an example where the KG lower bound is not optimal (Section~\ref{ssec:Cap_Erasure}).

\subsection{Secret Key Capacity for the WTC with Parallel Sources}
\label{ssec:Cap_SK}

We first analyze the secret key agreement problem for the model depicted in Fig.~\ref{fig:model_parallel}, where the nodes have access to correlated sources independent of the main channel. The upper bound for this model is found in Theorem~\ref{theo-KA-OB}, whereas the lower bound is derived from Theorem~\ref{cor-key} by taking the set of variables~\eqref{eq:parallel_var} and restricting the input probability distributions, cf.~\eqref{eq:theo-OB-pmf}, to the form: 
\begin{equation}
p(qu) p( x_c \vert u) p(y_c z_c \vert x_c) p(y_s \hat{y}_s z_s) p( t \vert v ) p( v \vert \hat{y}_s ) . \label{eq:ex_key_pmf}
\end{equation}
Then, the lower bound on the secret key rate~\eqref{eq:cor-key_rate} is given by
\begin{align}
\overline{\textnormal{C}}_{k} 
    &\geq \IC{V}{Y_s}{T} -\IC{V}{Z_s}{T} + \big| \I{U}{Y_c} -\IC{U}{Z_c}{Q} \nonumber\\
    &\quad -\max\{ \I{Q}{Y_c}, \, \IC{V}{\hat{Y}_s}{Y_s} \} \big|^+ , \label{eq:ex_key_rate}
\end{align}
maximized over~\eqref{eq:ex_key_pmf} and subject to
\begin{equation}
\IC{V}{\hat{Y}_s}{Y_s} \leq \I{U}{Y_c} . \label{eq:ex_key_cond}
\end{equation}
This bound is tight in some special cases.

\subsubsection{Eve Has a Less Noisy Channel}\label{sssec:ex_key_lnchann}

If Eve has a less noisy channel than Bob, no secrecy can be guaranteed in the main channel and the secret key is generated using only the correlated sources.

\vspace{1mm}
\begin{proposition}\label{prop:wsk}
In this scenario, the strong secret key capacity is given by
\begin{equation}
\overline{\textnormal{C}}_{k} =  \max_{p(x_c) p( t \vert v ) p( v \vert \hat{y}_s )}
 \big[ \IC{V}{Y_s}{T} -\IC{V}{Z_s}{T} \big] , \label{eq:ex_key_ln1_rate}
\end{equation}
subject to
\begin{equation}
\IC{V}{\hat{Y}_s}{Y_s} \leq \I{X_c}{Y_c} . \label{eq:ex_key_ln1_cond}
\end{equation}
\end{proposition}
\begin{IEEEproof}
For a given PD in~\eqref{eq:theo-OB-pmf} and given the less noisy condition on
Eve's channel, i.e., $\I{U}{Y_c} \leq \I{U}{Z_c}$ for any RV $U$ such that
$U\mkv X_c\mkv (Y_c Z_c)$, the upper bound from Theorem~\ref{theo-KA-OB} reduces
to~\eqref{eq:ex_key_ln1_rate}--\eqref{eq:ex_key_ln1_cond} which is equal to the
lower bound~\eqref{eq:ex_key_rate}--\eqref{eq:ex_key_cond} with $Q=\emptyset$
and $U=X_c$.
\end{IEEEproof}

\vspace{1mm}
\begin{remark}
The secret key capacity of the WTC with a public noiseless channel of rate $R$
\cite[Thm.~2.6]{csiszar_common_2000} is a special case of
Proposition~\ref{prop:wsk}, where $X_c = Y_c = Z_c$ and $\H{X_c}=R$.
This result was also noted in~\cite[Thm.~1]{prabhakaran_separation_2007}.
\end{remark}
\vspace{1mm}

\subsubsection{Eve Has a Less Noisy Side Information}

If Eve has a less noisy side information than Bob, the legitimate users cannot
extract any secret bits from the correlated sources; the key is the message
carried by the codeword $\V{U}$, which is secured from Eve by Wyner's
wiretap coding scheme.

\vspace{1mm}
\begin{proposition}\label{missing-prop2}
In this scenario, the strong secret key capacity is given by
\begin{equation}
\overline{\textnormal{C}}_{k} = \max_{p(u x_c)} \big[ \I{U}{Y_c} -\I{U}{Z_c} \big] . \label{eq:ex_key_ln2_rate}
\end{equation}
\end{proposition}
\begin{IEEEproof}
Given the less noisy condition on Eve's side information, i.e., $\I{V}{Y_s} \leq \I{V}{Z_s}$ for any RV $V$ such that $V\mkv \hat{Y}_s\mkv (Y_s Z_s)$, the upper bound reduces to~\eqref{eq:ex_key_ln2_rate} and the condition~\eqref{eq:theo_KA-OB-cond} disappears. Additionally, the lower bound~\eqref{eq:ex_key_rate}--\eqref{eq:ex_key_cond} achieves~\eqref{eq:ex_key_ln2_rate} with $Q=T=V=\emptyset$.
\end{IEEEproof}

\vspace{1mm}
\begin{remark}
Since the side information cannot be used to generate a secret key, the secret key capacity~\eqref{eq:ex_key_ln2_rate} is equal to the secrecy capacity of the WTC.
\end{remark}
\vspace{1mm}

\subsubsection{Alice and Bob Have the Same Side Information}
\label{sssec:ex_key_sameside}

If the legitimate users have access to the same side information, there is no need to transmit the bin indices of the description.

\vspace{1mm}
\begin{proposition}\label{missing-prop3}
In this scenario, the strong secret key capacity is given by
\begin{equation}
\overline{\textnormal{C}}_{k} =  \max_{p(u x_c)} \left[ \HC{Y_s}{Z_s} + \big| \I{U}{Y_c} -\I{U}{Z_c} \big|^+  \right] . \label{eq:ex_key_ln3_rate}
\end{equation}
\end{proposition}
\begin{IEEEproof}
If $\hat{Y}_s = Y_s$, the transmission cost of the description associated to the source disappears, i.e., $\IC{V}{\hat{Y}_s}{Y_s}=0$, which renders the conditions~\eqref{eq:theo_KA-OB-cond} and~\eqref{eq:ex_key_cond} redundant, and an achievable rate according to both the upper and lower bounds satisfies%
\begin{subequations}\label{eq:ex_key_ln3_1}
\begin{align}
R_k &\leq \IC{V}{Y_s}{Z_s} + \big| \I{U}{Y_c} -\I{U}{Z_c} \big|^+ \label{eq:ex_key_ln3_1c} \\
 &\leq \HC{Y_s}{Z_s} + \big| \I{U}{Y_c} -\I{U}{Z_c} \big|^+, \label{eq:ex_key_ln3_1d}
\end{align}
\end{subequations}
where
\begin{itemize}
 \item \eqref{eq:ex_key_ln3_1c} stems from the Markov chain $T\mkv V \mkv Y_s \mkv Z_s$ (due to $\hat{Y}_s = Y_s$), and $Q=\emptyset$ in the lower bound; and,
 \item in~\eqref{eq:ex_key_ln3_1d} we maximize the first term with $V=Y_s$. \hspace{1em plus 1fill}\IEEEQEDhere
\end{itemize}
\end{IEEEproof}

\subsection{Secrecy Capacity for the WTC with Parallel Sources}
\label{ssec:Cap_Secrecy}

We now study the secrecy capacity for the model depicted in Fig.~\ref{fig:model_parallel}. The upper bound for this model is found in Theorem~\ref{theo-OB}, whereas the lower bound can be derived from Theorem~\ref{theo-KG} by taking the set of variables~\eqref{eq:parallel_var} and restricting the input probability distributions to the form~\eqref{eq:ex_key_pmf}.
Then, the achievable secrecy rate $R_{KG_1}$~\eqref{eq:KG_region1_rate} is given by
\begin{subequations}\label{eq:ex_KG_region1_rate}
\begin{align}
 R_{KG_1} &\leq \IC{V}{Y_s}{T} -\IC{V}{Z_s}{T} +\I{U}{Y_c} -\IC{U}{Z_c}{Q} \nonumber\\
 &\quad -\max\{ \I{Q}{Y_c}, \, \IC{V}{\hat{Y}_s}{Y_s} \} , \\
 R_{KG_1} &\leq \I{U}{Y_c} -\max\{ \I{Q}{Y_c}, \, \IC{V}{\hat{Y}_s}{Y_s} \} ,
\end{align}
\end{subequations}
and $R_{KG_2}$~\eqref{eq:KG_region2_rate} by
\begin{subequations}\label{eq:ex_KG_region2_rate}
\begin{align}
R_{KG_2} &\leq \IC{V}{Y_s}{T} -\IC{V}{Z_s}{T} , \\
R_{KG_2} &\leq \I{U}{Y_c} -\IC{V}{\hat{Y}_s}{Y_s} .
\end{align}
\end{subequations}
This bound is tight in some special cases.

\subsubsection{Eve Has a Less Noisy Channel}

As in Section~\ref{sssec:ex_key_lnchann}, in the situation where Eve has a less noisy channel than Bob, the achievable secrecy rate is only due to the secret key generated using the correlated sources.

\vspace{1mm}
\begin{proposition}\label{missing-prop4}
In this scenario, the strong secrecy capacity is given by%
\begin{multline}
\overline{\textnormal{C}}_{s} =  \max_{p(x_c) p( t \vert v ) p( v \vert \hat{y}_s )}\, \min\! \Big\{ \IC{V}{Y_s}{T}
-\IC{V}{Z_s}{T} ,\\ \I{X_c}{Y_c} -\IC{V}{\hat{Y}_s}{Y_s} \Big\} .
\label{eq:ex_sec_ln1_rate}
\end{multline}
\end{proposition}
\begin{IEEEproof}
For a probability distribution in~\eqref{eq:theo-OB-pmf} and given the less noisy condition on
Eve's channel, the upper bound from Theorem~\ref{theo-OB} reduces
to~\eqref{eq:ex_sec_ln1_rate} which is equal to the lower
bound~\eqref{eq:ex_KG_region2_rate} with $U=X_c$.
\end{IEEEproof}
\vspace{1mm}

\subsubsection{Eve Has a Less Noisy Side Information}

If Eve has a less noisy side information than Bob, the legitimate users cannot
extract any secret bits from the correlated sources, and this problem reduces
to the wiretap channel.

\vspace{1mm}
\begin{proposition}
In this scenario, the strong secrecy capacity is given by
\begin{equation}
\overline{\textnormal{C}}_{s} = \max_{p(u x_c)} \big[ \I{U}{Y_c} -\I{U}{Z_c} \big] . \label{eq:ex_sec_ln2_rate}
\end{equation}
\end{proposition}
\begin{IEEEproof}
Given the less noisy condition on Eve's side information, the bound~\eqref{eq:theo_OB-rate1} becomes~\eqref{eq:ex_sec_ln2_rate} while the bound~\eqref{eq:theo_OB-rate2} becomes redundant. The bound~\eqref{eq:ex_sec_ln2_rate} is achieved by the lower bound~\eqref{eq:ex_KG_region1_rate} with $Q=T=V=\emptyset$.
\end{IEEEproof}
\vspace{1mm}

\subsubsection{Alice and Bob Have the Same Side Information and Bob Has a Less Noisy Channel}

Unlike Section~\ref{sssec:ex_key_sameside}, in order to achieve capacity the
legitimate users not only have to share the same side information but also Bob
needs a less noisy channel than Eve.

\vspace{1mm}
\begin{proposition}\label{prop:6}
In this scenario, the strong secrecy capacity is given by
\begin{multline}
\overline{\textnormal{C}}_{s} =  \max_{p(x_c)}\, \min \big\{ \I{X_c}{Y_c}, \\
\I{X_c}{Y_c} -\I{X_c}{Z_c} +\HC{Y_s}{Z_s} \big\} . \label{eq:ex_sec_ln3_rate}
\end{multline}
\end{proposition}
\begin{IEEEproof}
If $\hat{Y}_s = Y_s$, the transmission cost of the description associated to the source 
disappears, i.e., $\IC{V}{\hat{Y}_s}{Y_s}=0$, and following similar arguments as
those in~\eqref{eq:ex_key_ln3_1},
an achievable rate according to the upper bound~\eqref{eq:theo_OB-rate} satisfies
\begin{equation*}
R \leq \min \{ \I{X_c}{Y_c},\, \I{U}{Y_c} -\I{U}{Z_c} +\HC{Y_s}{Z_s} \} .
\end{equation*}
We may further upper-bound part of this expression as follows:%
\begin{align*}
\MoveEqLeft[1]
\I{U}{Y_c} -\I{U}{Z_c} \\
 &= \I{X_c}{Y_c} -\IC{X_c}{Y_c}{U} -\I{X_c}{Z_c} +\IC{X_c}{Z_c}{U} \\
 &\leq \I{X_c}{Y_c} -\I{X_c}{Z_c} ,
\end{align*}
where the inequality is due to Bob's channel being less noisy than Eve's. 
Hence, the upper bound becomes~\eqref{eq:ex_sec_ln3_rate} under the aforementioned
conditions, which is achieved by the lower bound~\eqref{eq:ex_KG_region1_rate}
with $Q=T=\emptyset$, $V=Y_s$, and $U=X_c$.
\end{IEEEproof}

\subsection{Wiretap Channel with Perfect Output Feedback}
\label{ssec:Cap_WTC-POF}

In~\cite{ahlswede_trans_2006}, the authors analyze a wiretap channel with
perfect output feedback at the encoder, i.e., $\hat{Y} = Y$, and perfectly
secured from the eavesdropper.

\vspace{1mm}
\begin{theorem}[{\hspace{1sp}\cite[Thm. 1]{ahlswede_trans_2006}}]\label{theorem-5}
In this model, the KG lower bound introduced in Theorem~\ref{theo-KG} achieves all
rates satisfying
\begin{multline}
 R \leq \max\limits_{p(ux)}\, \min \big\{ \I{U}{Y} , \\
 | \I{U}{Y} -\I{U}{Z} |^+ +\HC{Y}{UZ}\big\} . \label{eq:perfect_rate}
\end{multline}
\end{theorem}
\begin{IEEEproof}
With the following choice of RVs
\begin{align*}
 V &= Y &
 &\textnormal{ and }&
 T &= Q = \emptyset ,
\end{align*}
the achievable secrecy rate $R_{KG_1}$~\eqref{eq:KG_region1_rate} becomes
\begin{equation*}
 R_{KG_1} \leq \min \{ \I{U}{Y} -\I{U}{Z} +\HC{Y}{UZ}, \, \I{U}{Y} \} ,
\end{equation*}
while the achievable secrecy rate $R_{KG_2}$~\eqref{eq:KG_region2_rate} reads 
\begin{equation*}
 R_{KG_2} \leq \min \{ \HC{Y}{UZ},\, \I{U}{Y} \} .
\end{equation*}
Therefore, the maximization over both strategies can be succinctly written as~\eqref{eq:perfect_rate}.
\end{IEEEproof}

\vspace{1mm}
\begin{remark}
The secrecy capacity results for the \emph{degraded} and \emph{reversely degraded} WTC with perfect output feedback~\cite[Cor. 1 and 2]{ahlswede_trans_2006} also apply here.
\end{remark}

\subsection{Wiretap Channel with Causal State Information}
\label{ssec:Cap_WTC-CSI}

In~\cite{chia_wiretap_2012}, the authors analyze a wiretap channel affected
by a random state $S$, i.e., $p(yz\vert xs)p(s)$, where the state is available
causally only at the encoder and the legitimate decoder, i.e., $\hat{Y} = S$
and $Y = (Y,S)$.

\vspace{1mm}
\begin{theorem}[{\hspace{1sp}\cite[Thm. 1]{chia_wiretap_2012}}]\label{theorem-6}
In this model, a slightly modified version of the KG scheme presented in Theorem~\ref{theo-KG} achieves all the rates satisfying
\begin{align}
 R \leq \max \bigg\{
    \max_{p(u)u'(u,s)p(x\vert u's)}
	&\min \{ \I{U}{YS} -\I{U}{ZS} \nonumber\\ &\qquad+\HC{S}{Z}, \, \I{U}{YS}\}, \nonumber\\
    \max_{p(u)p(x\vert us)} \quad
	&\min \{ \HC{S}{ZU}, \, \IC{U}{Y}{S}\} \bigg\} . \label{eq:state_rate}
\end{align}
\end{theorem}
\begin{IEEEproof}
First, we make the choice of RVs
\begin{align*}
 V &= S &
 &\textnormal{ and }&
 T &= Q = \emptyset .
\end{align*}
Second, since the KG scheme is derived to handle \emph{strictly} causal feedback, and the present model assumes the state is known causally at the encoder, i.e., $s^i$ is present at time slot $i$, we need to perform a slight modification of the scheme.

We can modify step~\ref{it:KG_encod_3}) from the encoding procedure (Appendix~\ref{ssec:KG_enc}) in the following way. For $R_{KG_1}$,
after the encoder has chosen the codeword to transmit in block $j$, i.e.,
$\V{u}(\ve{r}_j)$, it computes $u_i'=u'(u_i(\ve{r}_j), s_i)$ and transmits a
randomly generated symbol $x_i$ according to $p(x_i\vert u_i' s_i)$ for each
time slot $i\in[1:n]$. The rate~\eqref{eq:KG_region1_rate} becomes%
\begin{align*}
 R_{KG_1} &\leq \I{U}{YS} -\I{U}{Z} +\HC{S}{ZU} \nonumber\\
          &=    \I{U}{YS} -\I{U}{ZS}+\HC{S}{Z} , \\
 R_{KG_1} &\leq \I{U}{YS} .
\end{align*}
For $R_{KG_2}$, we proceed similarly but without the inclusion of the function
$u'(\cdot)$ between the codeword $\V{u}(\ve{r}_j)$ and the generation of $x_i$.
The rate~\eqref{eq:KG_region2_rate} becomes
\begin{align*}
 R_{KG_2} &\leq \IC{S}{YS}{U} -\IC{S}{Z}{U} = \HC{S}{ZU} , \\
 R_{KG_2} &\leq \I{U}{YS} = \IC{U}{Y}{S} .
\end{align*}

Therefore, the final expression for the rate is~\eqref{eq:state_rate}.
\end{IEEEproof}

\vspace{1mm}
\begin{remark}
The secrecy capacity result for \emph{less noisy} WTC with state information
available causally or noncausally at the encoder and decoder~\cite[Thm.
3]{ahlswede_trans_2006} also applies here.
\end{remark}

\begin{figure*}[t!]
\centering
\begin{tikzpicture}[line width=.6pt, font=\footnotesize]
\draw (1.375,3) node {$\V{q}(l')$};	
\fill[black!35!white] (1,2) rectangle (1.75,2.25);
\foreach \i in {0,...,6} { %
    \draw[very thin] (1,1+\i*0.25) rectangle (1.75,1.25+\i*0.25);
}
\draw[thick] (1,1) rectangle (1.75,2.75);
\draw[black!25!white, dashed] (1,2.25) -- (4.75,4.25);
\draw[black!25!white, dashed] (1.75,2) -- (5.5,3);
\draw[black!25!white, dashed] (1,2) -- (4,.75);
\draw[black!25!white, dashed] (1.75,2.25) -- (6.25,1.75);
\draw[<->] (.75,2.75) -- (.75,1) node[midway,xshift=-11pt] {$2^{n\tilde{S}'}$};
%
\draw (5.125,4.5) node {$\V{u}(\ve{r})$};
\fill[black!35!white] (4.75,3.75) rectangle (5.5,4);
\foreach \i in {0,...,4} { %
    \draw[very thin] (4.75,3+\i*0.25) rectangle (5.5,3.25+\i*0.25); 
}
\draw[thick] (4.75,3) rectangle (5.5,4.25);
\draw[black!25!white, dashed] (4.75,3.75) -- (7,1.5);
\draw[black!25!white, dashed] (5.5,4) -- (13.75,3.25);
\draw[<->] (4.5,4.25) -- (4.5,3) node[midway,xshift=-37pt] {$2^{n(\tilde{S}''+R_0+R_1+R_f)}$};
%
\draw (5.125,2) node {$\V{t}(l',s_1)$};
\fill[black!35!white] (5.5,1) rectangle (6.25,1.25);
\foreach \i in {0,...,3} { %
  \foreach \j in {0,...,2} { %
    \draw[very thin] (4+\j*.75,.75+\i*0.25) rectangle (4.75+\j*.75,1+\i*0.25); 
  }
}
\foreach \j in {0,...,2} { %
  \draw[thick] (4+\j*.75,.75) rectangle (4.75+\j*.75,1.75); 
}
\draw[thick] (4,.75) rectangle (5.5,1.75);
\draw[black!25!white, dashed] (5.5,1.25) -- (7,3.25);
\draw[black!25!white, dashed] (6.25,1) -- (13.75,1.5);
\draw[<->] (3.75,.75) -- (3.75,1.75) node[midway,xshift=-20pt]
{$2^{n(S_1-\tilde{S}_1)}$};
\draw [decorate,decoration={brace,amplitude=5pt}] (4.75,.6) -- (4,.6) node
[midway,yshift=-12pt] {$B_1(1)$};
\draw [decorate,decoration={brace,amplitude=5pt}] (6.25,.6) -- (5.5,.6) node
[midway,yshift=-11pt] {$B_1(2^{n\tilde{S}_1})$};
%
\draw (10.375,3.5) node {$\V{v}(\ve{r},s_1,s_2)$};
\foreach \i in {0,...,6} { %
  \foreach \j in {0,...,8} { %
    \draw[very thin] (7+\j*.75,1.5+\i*0.25) rectangle (7.75+\j*.75,1.75+\i*0.25); 
  }
}
\foreach \j in {0,...,2} { %
  \draw[thick] (7+\j*2.25,1.5) rectangle (9.25+\j*2.25,3.25); 
}
\draw[thick] (7,1.5) rectangle (13.75,3.25);
\draw[<->] (14,3.25) -- (14,1.5) node[midway,right] {$2^{n(S_2-\tilde{S}_2-\bar{S}_2)}$};
\draw [decorate,decoration={brace,amplitude=5pt}] (9.25,1.35) -- (7,1.35) node [midway,yshift=-12pt] {$B_2(1)$};
\draw [decorate,decoration={brace,amplitude=5pt}] (13.75,1.35) -- (11.5,1.35) node [midway,yshift=-11pt] {$B_2(2^{n\tilde{S}_2})$};
\draw [decorate,decoration={brace,amplitude=4pt}] (10,1.4) -- (9.25,1.4) node [midway,yshift=-11pt] {\scriptsize $\bar{B}_2(l_2,1)$};
%
\draw [decorate,decoration={brace,amplitude=4pt}] (11.5,1.4) -- (10.75,1.4) node [midway,yshift=-10pt] {\scriptsize $\bar{B}_2(l_2,2^{n\bar{S}_2})$};
\end{tikzpicture}
\caption{Schematic representation of the codebook. The index $s_1$ in the bins and sub-bins of $\V{v}(\cdot)$ is not shown to improve readability.
\label{fig:KG_codebooks}}
\end{figure*}
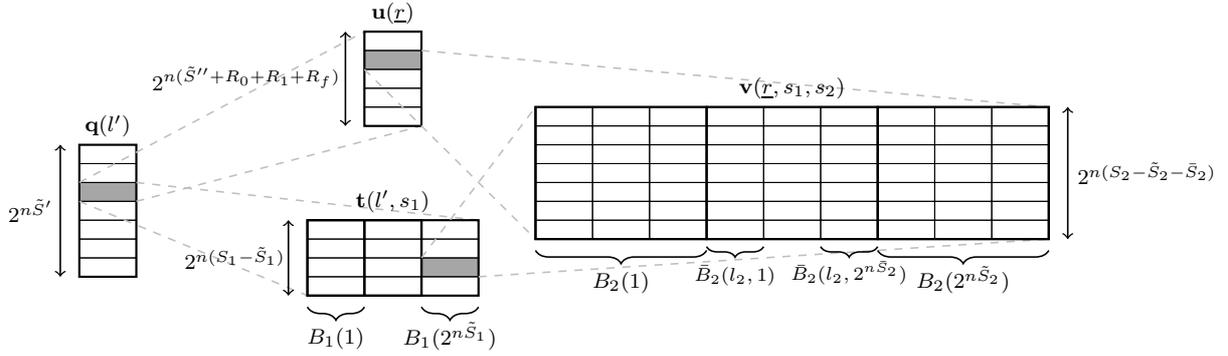

\subsection{Erasure Wiretap Channel with State-Feedback}
\label{ssec:Cap_Erasure}

In~\cite{czap_secret_2015}, the authors analyze the erasure WTC with public state-feedback from the legitimate receiver; therefore, both the encoder and the eavesdropper know if there was an erasure or not at the legitimate end. 
In other words, let $S\triangleq\mathds{1}\{ Y=e \}$ indicate the erasure event at the legitimate user, then
\begin{align}
 \hat{Y} &\triangleq S &
 &\textnormal{and}&
 Z &\triangleq (Z', S) , \label{eq:erasure_rv}
\end{align}
where $Z'$ is the eavesdropper's channel output.
Moreover, the channels experience independent erasures, i.e., $p(yz'\vert x)= p(y\vert x)p(z'\vert x)$.

\vspace{1mm}
\begin{proposition}\label{prop:erasure_kg}
In this scenario, it can be shown that the KG lower bound from Theorem~\ref{theo-KG} achieves any rate
\begin{equation}
 R \leq (1-\delta) \delta_E \max \left\{ \frac{\!1 -\delta}{1 -\delta\delta_E}, \, \frac{1}{1 +\delta_E} \right\} , \label{eq:erasure_kg}
\end{equation}
where $\delta$ denotes the erasure probability of the legitimate receiver and $\delta_E$, the one of the eavesdropper. 
\end{proposition}
\begin{IEEEproof}
 See Appendix~\ref{sec:Proof-Prop-Erasure}.
\end{IEEEproof}
\vspace{1mm}

Even though~\eqref{eq:erasure_kg} is the maximum secrecy rate achieved by the KG scheme, it is strictly suboptimal.
The secrecy capacity of this channel model is given by~\cite[Cor. 1]{czap_secret_2015}
\begin{equation}
 \overline{\textnormal{C}}_{sf} = (1-\delta) \delta_E \frac{1-\delta \delta_E}{1- \delta \delta_E^2} , \label{eq:erasure_cap}
\end{equation}
and numerical analysis shows that~\eqref{eq:erasure_kg} is strictly below~\eqref{eq:erasure_cap} for all $\delta$ and $\delta_E\in (0,1)$.

\section{Summary and Concluding Remarks}
\label{sec:Summary}

In this work, we presented an achievable scheme for the wiretap channel with generalized feedback, the KG lower bound, which allows the legitimate users to agree on a secret key simultaneously with the transmission of a message.
As an extension to this scheme, we introduced a strategy for the problem of secret key agreement, which is essentially the KG lower bound when no message is transmitted.

Due to the complexity of the general problem, we resorted to simpler channel models to characterize the merit of these schemes. For a special class of channels, which we named wiretap channel with parallel sources, we derived two novel upper bounds and we showed the optimality of the KG lower bound and its secret key counterpart under some special conditions.
As a side note, it should be mentioned that the capacity result in Proposition~\ref{missing-prop4} was recently re-discovered in~\cite[Cor.~1]{goldfeld_states_2016} by employing a different coding scheme than our work in~\cite{bassi_generalized_2015}.

In addition to these new capacity results, the KG lower bound also recovered previously reported results for different channel and feedback models.
Consequently, this lower bound could be seen as a generalization, and hence unification of several results in the field.
Nonetheless, the unification is not complete since the KG lower bound failed to recover all known results, as shown in Section~\ref{ssec:Cap_Erasure}. 


\appendices

\section{Proof of Theorem~\ref{theo-KG} (KG Lower Bound)}
\label{sec:Proof_KG}

The encoder splits the transmission in $b$ blocks of $n$ channel uses, during which it transmits $b-1$ messages of rate $R$.
During each transmission block and in addition to the messages, the encoder also sends the bin indices corresponding to two layers of description of the feedback sequence it observed in the previous block. This allows the legitimate users to agree on a secret key which is used to encrypt part of the transmission.

The messages are sent using one of the following two strategies.
In the first one, the rate $R=R_{KG_1}$ is achievable by the joint use of Wyner's wiretap coding scheme, which provides a secure rate of $R_0$ bits, and a bitwise-encrypted message, which grants the remaining $R_1=R-R_0$ secure bits. 
The second strategy only relies on the aforementioned secret key to send an encrypted message of rate $R=R_{KG_2}$. 

In the sequel, we present the proof for $R_{KG_1}$ in detail while only a sketch of the proof of $R_{KG_2}$ is provided after that.
We note that the rates are shown to be achievable according to the weak secrecy condition~\eqref{eq:s_rate_weak}. Nonetheless, we demonstrate at the end of this Appendix that the strong secrecy condition~\eqref{eq:s_rate_strong} also holds true.

\subsection{Codebook Generation}\label{ssec:KG_code}

Let us define the quantities
\begin{subequations}\label{eq:KG_rates}
\begin{align}
S_1         &= \IC{T}{UX\hat{Y}}{Q} +\epsilon_1 , \\
\tilde{S}_1 &= \IC{T}{UX\hat{Y}}{Q} -\IC{T}{UY}{Q} +\epsilon_1 +\tilde{\epsilon}_1 , \\
S_2         &= \IC{V}{X\hat{Y}}{UT} +\epsilon_2 , \\
\tilde{S}_2 &= \IC{V}{X\hat{Y}}{UT} -\IC{V}{Y}{UT} +\epsilon_2 +\tilde{\epsilon}_2 \mathrlap{,} \displaybreak[2]\\
\bar{S}_2   &= \IC{V}{Y}{UT} -\IC{V}{Z}{UT} , \\
R_1 +R_f    &= \IC{U}{TZ}{Q} -\epsilon' , \label{eq:KG_rates_6}
\end{align}
\end{subequations}
and fix the joint distribution~\eqref{eq:KG_pmf} that achieves the maximum
in $R_{KG_1}$. Then, for each block, create independent codebooks as follows:
\begin{enumerate}
 \item Randomly pick $2^{n\tilde{S}'}$ sequences $\V{q}(l')$, $l' \in [1:2^{n\tilde{S}'}]$, from $\typ{Q}$.
 \item \label{it:kg-u} For each $\V{q}(l')$, randomly pick $2^{n(\tilde{S}''+R_0+R_1+R_f)}$ sequences $\V{u}(\ve{r}) \equiv \V{u}(l',l'',m_0,m_1,l_f)$, where $l'' \in [1:2^{n\tilde{S}''}]$, $m_0 \in [1:2^{nR_0}]$, $m_1 \in [1:2^{nR_1}]$, and $l_f \in [1:2^{nR_f}]$, from $\typ{U|\V{q}(l')}$.
 \item For each $\V{q}(l')$, randomly pick $2^{nS_1}$ sequences $\V{t}(l',s_1)$, where $s_1 \in [1:2^{nS_1}]$, from $\typ{T|\V{q}(l')}$.
 Distribute the sequences uniformly at random in $2^{n\tilde{S}_1}$ equal-sized bins $B_1(l_1)$, which is possible since $\tilde{S}_1 \leq S_1$.
 \item For each possible triplet $(\V{q}(l'), \V{u}(\ve{r}), \V{t}(l',s_1))$, randomly pick $2^{nS_2}$ sequences $\V{v}(\ve{r},s_1,s_2)$, where $s_2 \in [1:2^{nS_2}]$, from $\typ{V|\V{q}(l'), \V{u}(\ve{r}), \V{t}(l',s_1)}$.
 Distribute the sequences uniformly at random in $2^{n\tilde{S}_2}$ equal-sized bins $B_2(s_1,l_2)$ and the sequences in each bin in $2^{n\bar{S}_2}$  equal-sized sub-bins $\bar{B}_2(s_1,l_2,k)$. This binning process is feasible if
 \begin{subequations}
 \begin{align}
  \tilde{S}_2 &\leq S_2 , \\
  \bar{S}_2   &\leq S_2 -\tilde{S}_2 ,
 \end{align}
 \end{subequations}
 which holds according to~\eqref{eq:KG_rates} as long as $\IC{V}{Z}{UT} \leq \IC{V}{Y}{UT}$.
 Moreover, partition the set $[1:2^{n\bar{S}_2}]$ in $2^{nR_1}$ equal-sized subsets, which defines the mapping $k' = M_k(k)$, where $k' \in 
[1:2^{nR_1}]$. This partition is possible if
 \begin{equation}
  R_1 \leq \bar{S}_2 .
 \end{equation}
\end{enumerate}


See Fig.~\ref{fig:KG_codebooks} for details.

\subsection{Encoding}\label{ssec:KG_enc}

In the first block, the encoder chooses a codeword $\V{u}(\ve{r}_1)$ uniformly at random.
It then transmits the sequence $\V{x}_1$ that
is randomly generated according to the conditional PD $p(\V{x}\vert \V{u}(\ve{r}_1)) =
\prod_{i=1}^n p(x_i \vert u_i(\ve{r}_1))$.

\vspace{1mm}
In block $j \in [2:b]$, proceed as follows:
\begin{enumerate}
 \item Given the channel input and the feedback signal from the previous block, 
the encoder looks for an index $s_{1(j-1)}\equiv \hat{s}_1$ such that
\begin{multline*}
 \left( \V{t}(l_{j-1}',\hat{s}_1), \V{q}(l_{j-1}'), \V{u}(\ve{r}_{j-1}), \V{x}_{j-1}, \V{\hat{y}}_{j-1} \right) \\
 \in \typc{TQUX\hat{Y}} ,
\end{multline*}
where $\delta'<\epsilon_1$.
If more than one index is found, choose one uniformly at random, whereas if there is no such index, choose one uniformly at random in $[1:2^{nS_1}]$.
The probability of not finding such an index is arbitrarily small as $n\to\infty$.
 \item Then, the encoder looks for an index $s_{2(j-1)}\equiv \hat{s}_2$ such that
\begin{multline*}
 \big( \V{v}(\ve{r}_{j-1},s_{1(j-1)},\hat{s}_2), \V{t}(l_{j-1}',s_{1(j-1)}), \V{q}(l_{j-1}'), \\
 \V{u}(\ve{r}_{j-1}), \V{x}_{j-1}, \V{\hat{y}}_{j-1} \big) \in \typc{VTQUX\hat{Y}} ,
\end{multline*}
where $\delta'<\epsilon_2$.
If more than one index is found, choose one uniformly at random, whereas if there is no such index, choose one uniformly at random in $[1:2^{nS_2}]$.
The probability of not finding such an index is arbitrarily small as $n\to\infty$.
 \item Let $\V{v}(\ve{r}_{j-1}, s_{1(j-1)},s_{2(j-1)}) \mkern-2mu \in \mkern-2mu \bar{B}_2(s_{1(j-1)},l_{2(j-1)},k_{j-1})$ and 
 $\V{t}(l_{j-1}',s_{1(j-1)}) \in B_1(l_{1(j-1)})$, 
 and define the following mapping. Let 
 $(l_{j}', l_{j}'') = M_l(l_{1(j-1)},l_{2(j-1)})$, such that $M_l(\cdot)$ is invertible. This function can be defined if
 \begin{equation}
  \tilde{S}' +\tilde{S}'' = \tilde{S}_1 +\tilde{S}_2 . 
 \end{equation}
 \item \label{it:KG_encod_3}
In order to transmit the message $m_j=(m_{0j}, m_{1j})$, the encoder 
chooses uniformly at random a value for the index $l_{fj} \in [1:2^{nR_f}]$ and selects the 
codeword $\V{u}(l_{j}', l_{j}'', m_{0j}, m_{1j}',l_{fj}) = \V{u}(\ve{r}_j)$,
where $m_{1j}' = m_{1j} \oplus k_{j-1}'$ and $k_{j-1}'=M_k(k_{j-1})$. It then transmits 
the sequence $\V{x}_j$ that is randomly generated according to the conditional PD $p(\V{x}\vert \V{u}(\ve{r}_j)) =
\prod_{i=1}^n p(x_i \vert u_i(\ve{r}_j))$.
\end{enumerate}

\subsection{Decoding}\label{ssec:KG_dec}

At the end of each transmission block $j \in [1:b]$, the legitimate decoder looks for the unique set of indices $\ve{r}_j = (l_{j}', l_{j}'', m_{0j}, m_{1j}',l_{fj}) \equiv (\hat{l}', \hat{l}'', \hat{m}_0, \hat{m}_1', \hat{l}_f)$ such that
\begin{equation*}
 \big( \V{q}(\hat{l}'), \V{u}(\hat{l}', \hat{l}'', \hat{m}_0, \hat{m}_1', \hat{l}_f), \V{y}_j \big) \in \typ{QUY} .
\end{equation*}
 The probability of error in decoding can be made arbitrarily small provided that
\begin{subequations}
\begin{align}
 \tilde{S}'' +R_0 +R_1 +R_f &< \IC{U}{Y}{Q} -\delta , \\
 \tilde{S}' +\tilde{S}'' +R_0 +R_1 +R_f &< \I{U}{Y} -\delta .
\end{align}
\end{subequations}

Additionally, in block $j \in [2:b]$, proceed as follows:
\begin{enumerate}
%
%
 \item The legitimate decoder computes $(l_{1(j-1)},l_{2(j-1)}) = M_l^{-1}(l_{j}', l_{j}'')$.
 \item It then looks for the unique index $s_{1(j-1)} \equiv \hat{s}_1$ such that $\V{t}(l_{j-1}', \hat{s}_1) \in B_1(l_{1(j-1)})$ and
\begin{equation*}
 \left( \V{t}(l_{j-1}', \hat{s}_1), \V{q}(l_{j-1}'), \V{u}(\ve{r}_{j-1}), \V{y}_{j-1} \right) \in \typ{TQUY} ,
\end{equation*}
 where $\delta<\tilde{\epsilon}_1$. The probability of error in decoding is arbitrarily small as $n\to\infty$.
 \item \label{it:dec_v} The legitimate decoder additionally looks for the unique index $s_{2(j-1)} \equiv \hat{s}_2$ such that $\V{v}(\ve{r}_{j-1}, s_{1(j-1)}, \hat{s}_2) \in B_2(s_{1(j-1)},l_{2(j-1)})$ and
\begin{multline*}
 \big( \V{v}(\ve{r}_{j-1}, s_{1(j-1)}, \hat{s}_2), \V{t}(l_{j-1}', s_{1(j-1)}), \V{q}(l_{j-1}'), \\
 \V{u}(\ve{r}_{j-1}), \V{y}_{j-1} \big) \in \typ{VTQUY} ,
\end{multline*}
 where $\delta<\tilde{\epsilon}_2$. The probability of error in decoding is arbitrarily small as $n\to\infty$.
 \item The legitimate decoder is therefore able to recover the secret key $k_{j-1}' = M_k(k_{j-1})$ 
from the sub-bin $k_{j-1}$, i.e., $\V{v}(\ve{r}_{j-1}, s_{1(j-1)},s_{2(j-1)}) \in 
\bar{B}_2(s_{1(j-1)},l_{2(j-1)},k_{j-1})$, and with this key, it decrypts the message of the present block,
i.e., $m_j=(m_{0j}, m_{1j}'\oplus k_{j-1}')$.
\end{enumerate}

\subsection{Key Leakage}

Let us denote with $L_{1j}$ the random variable associated with the bin index of codeword $\V{T}_j$ in block $j$, and $L_{2j}$ and $K_j$ the random variables associated with the bin and sub-bin index of codeword $\V{V}_j$ in block $j$, respectively.

\vspace{1mm}
\begin{remark}\label{rk:KG-markov}
Owing to the encoding procedure, the variables $L_{1j}$, $L_{2j}$, and $K_j'=M_k(K_j)$ are the only cause of the correlation between blocks, the latter through $\bM_{1(j+1)}'=\bM_{1(j+1)}\oplus K_{j}'$. This fact is used in many of the subsequent Markov chains.
\end{remark}
\vspace{1mm}

Consider the following,
\begin{subequations}\label{eq:KG_keyleak_1}
\begin{align}
\MoveEqLeft[1]
\HC{K^{b-1}}{\cC \V{Z}^b} & \nonumber\\
 &= \sum\nolimits_{j=1}^{b-1} \HC{K_j}{\cC \V{Z}^b K^{j-1}} \nonumber\\
%
%
 &\geq \sum\nolimits_{j=1}^{b-1} \HC{K_j}{\cC \V{U}_j \V{Z}_j^b} \label{eq:KG_keyleak_1a} \displaybreak[2]\\
 &\geq \sum\nolimits_{j=1}^{b-1} \HC{K_j}{\cC \V{U}_j \V{Z}_j L_{1j} L_{2j} \bM_{1(j+1)}'} \label{eq:KG_keyleak_1b} \displaybreak[2]\\
 &\geq \sum\nolimits_{j=1}^{b-1} \HC{K_j}{\cC \V{U}_j \V{Z}_j \V{T}_j L_{2j} \bM_{1(j+1)}'} \nonumber\displaybreak[2]\\
 &= \sum\nolimits_{j=1}^{b-1} \HC{K_j \V{X}_j \V{\hat{Y}}_j}{\cC \V{U}_j \V{Z}_j \V{T}_j L_{2j} \bM_{1(j+1)}'} \nonumber\\
 &\qquad\qquad -\HC{\V{X}_j \V{\hat{Y}}_j}{\cC \V{U}_j \V{Z}_j \V{T}_j L_{2j} K_j}, \label{eq:KG_keyleak_1c}
\end{align}
\end{subequations}
where
\begin{itemize}
 \item \eqref{eq:KG_keyleak_1a} is due to $(\V{Z}^{j-1} K^{j-1}) \mkv (\cC \V{U}_j) \mkv (\V{Z}_j^b K_j)$ being a Markov chain since $\V{U}_j$ contains $(L_{1(j-1)} L_{2(j-1)} K_{j-1}')$, see Remark~\ref{rk:KG-markov}; and,
 \item \eqref{eq:KG_keyleak_1b} is due to $\V{Z}_{j+1}^b \mkv (\cC L_{1j} L_{2j} \bM_{1(j+1)}') \mkv (K_j \V{U}_j \V{Z}_j)$.
\end{itemize}
The first term in~\eqref{eq:KG_keyleak_1c} can be bounded from below as follows,%
\begin{subequations}\label{eq:KG_keyleak_2}
\begin{align}
\MoveEqLeft[1]
\HC{\V{X}_j \V{\hat{Y}}_j}{\cC \V{U}_j \V{T}_j \V{Z}_j L_{2j} \bM_{1(j+1)}'}  \nonumber\\
 &= \HC{\V{X}_j \V{\hat{Y}}_j}{\cC \V{U}_j \V{T}_j \V{Z}_j} 
   -\IC{\V{X}_j \V{\hat{Y}}_j}{L_{2j}}{\cC \V{U}_j \V{T}_j \V{Z}_j} \nonumber\\
 &\quad-\IC{\V{X}_j \V{\hat{Y}}_j}{\bM_{1(j+1)}'}{\cC \V{U}_j \V{T}_j \V{Z}_j L_{2j}} \nonumber\\
 &\geq \HC{\V{X} \V{\hat{Y}}}{\cC \V{U} \V{T} \V{Z}} -\H{L_{2j}} -\I{K_{j}'}{\bM_{1(j+1)}'} \label{eq:KG_keyleak_2a}\\
 &\geq \HC{\V{X} \V{\hat{Y}}}{\cC \V{U} \V{T} \V{Z}} -n\tilde{S}_2 \label{eq:KG_keyleak_2b} \displaybreak[2]\\
 &\geq \HC{\V{X} \V{\hat{Y}} \V{T} \V{Z}}{\cC \V{U}} -\HC{\V{T} \V{Z}}{\cC \V{U}} -n\tilde{S}_2 \nonumber \displaybreak[2]\\
 &\geq \HC{\V{X} \V{\hat{Y}} \V{Z}}{\cC \V{U}} -\HC{\V{Z}}{\cC \V{U}} -\HC{\V{T}}{\cC \V{U} \V{Z}} -n\tilde{S}_2 \nonumber \displaybreak[2]\\
 &\geq n \big[ \HC{X\hat{Y}}{U Z} -\epsilon' \big] -\HC{\V{T}}{\cC \V{U} \V{Z}} -n\tilde{S}_2 \label{eq:KG_keyleak_2c} \\
 &\geq n \big[ \HC{X\hat{Y}}{U T Z} -\epsilon_1 -\eta -\epsilon' -\tilde{S}_2 \big] , \label{eq:KG_keyleak_2d}
\end{align}
\end{subequations}
where
\begin{itemize}
 \item \eqref{eq:KG_keyleak_2a} is due to $\bM_{1(j+1)}' \mkv K_j' \mkv (\cC \V{U}_j \V{T}_j \V{X}_j \V{\hat{Y}}_j \V{Z}_j L_{2j})$ being a Markov chain, and the block index $j$ in the first term being removed for notational simplicity;
 \item \eqref{eq:KG_keyleak_2b} is due to $\H{L_{2j}}\leq n\tilde{S}_2$, and $\H{\bM_{1(j+1)}\oplus K_{j}'}=\H{\bM_{1(j+1)}}$ since $\bM_{1(j+1)}$ is uniformly distributed on $[1:2^{nR_1}]$ and independent of $K_{j}'$;
 \item \eqref{eq:KG_keyleak_2c} is due to $\cC\mkv \V{U}\mkv (\V{X} \V{\hat{Y}} \V{Z})$ being a Markov chain, and $\HC{\V{X} \V{\hat{Y}} \V{Z}}{\V{U}} \geq n \big[ \HC{X\hat{Y}Z}{U} -\epsilon' \big]$ for some $\epsilon'>0$ since all the sequences are jointly typical%
 \footnote{Given the encoding procedure, \V{X} is generated in an i.i.d. fashion given \V{U}, and thus $p(\V{x} \V{\hat{y}} \V{z}\vert \V{u}) = \prod_i p(x_i \hat{y}_i z_i\vert u_i)$.
 Although it is not true in general that $p(\V{\hat{y}} \V{z}\vert \V{x}) = \prod_i p(\hat{y}_i z_i\vert x_i)$ due to the use of feedback in the encoding procedure, cf.~\eqref{eq:wtc_nth_extension}, the scheme only correlates \emph{adjacent} transmission blocks. Therefore, \emph{inside} a transmission block, we have a DMC without feedback.}; and,
 \item \eqref{eq:KG_keyleak_2d} stems from the following lemma\footnote{Although \V{Q} is not explicitly denoted in the conditioning of the entropy in~\eqref{eq:KG_keyleak_2c}, it is assumed to be there hidden behind \V{U}.}.
\end{itemize}

\vspace{1mm}
\begin{lemma}\label{lm:KG-Bound}
Let $\eta>0$ and $\epsilon_1$ defined in~\eqref{eq:KG_rates}. Then, given the codebook generation and encoding procedure of the scheme,
\begin{equation}
\HC{\V{T}}{\cC \V{Q} \V{U} \V{Z}} \leq n \big[ \IC{T}{X\hat{Y}}{UZ} +\epsilon_1 +\eta \big] ,
\end{equation}
for sufficiently large $n$.
\end{lemma}
\begin{IEEEproof}
The proof is found in Appendix~\ref{sec:Proof-Lemma-Bound}.
\end{IEEEproof}
\vspace{1mm}

On the other hand, the second term in~\eqref{eq:KG_keyleak_1c} can be bounded from above as
\begin{align}
\MoveEqLeft[1]
\HC{\V{X} \V{\hat{Y}}}{\cC \V{U} \V{Z} \V{T} L_2 K} \nonumber\\
 &= \HC{\V{X} \V{\hat{Y}}}{\cC \V{U} \V{Z} \V{T} \V{V}} +\IC{\V{X} \V{\hat{Y}}}{\V{V}}{\cC \V{U} \V{Z} \V{T} L_2 K} \nonumber\\
 &\leq n\HC{X\hat{Y}}{UTVZ} +\HC{\V{V}}{\cC \V{U} \V{Z} \V{T} L_2 K} \nonumber\\
 &\leq n \big[ \HC{X\hat{Y}}{UTVZ} +\epsilon_n \big], \label{eq:KG_keyleak_3}
\end{align}
where the last inequality stems from the following lemma.

\vspace{1mm}
\begin{lemma}\label{lm:KG-Fano}
Given the codebook generation and encoding procedure of the scheme,
\begin{equation}
\HC{\V{V}}{\cC \V{U} \V{Z} \V{T} L_2 K} \leq n\epsilon_n , \label{eq:lemma_fano}
\end{equation}
where $\epsilon_n$ denotes a sequence such that $\epsilon_n\to 0$ as $n\to\infty$.
\end{lemma}
\begin{IEEEproof}
The proof is found in Appendix~\ref{sec:Proof-Lemma-Fano}.
\end{IEEEproof}
\vspace{1mm}

Therefore, joining~\eqref{eq:KG_keyleak_1}, \eqref{eq:KG_keyleak_2}, and~\eqref{eq:KG_keyleak_3}, we obtain
\begin{align}
\MoveEqLeft[1]
\HC{K^{b-1}}{\cC \V{Z}^b} & \nonumber\\
 &\geq \sum\nolimits_{j=1}^{b-1} n \big[ \IC{V}{X\hat{Y}}{U T Z} -\tilde{S}_2 -\epsilon_1 -\eta -\epsilon_n \big] \nonumber\\
 &= \sum\nolimits_{j=1}^{b-1} n \big[ \bar{S}_2 -(\epsilon_1 +\epsilon_2 +\tilde{\epsilon}_2 +\eta +\epsilon_n) \big] \nonumber\\
 &= n (b-1) (\bar{S}_2 -\epsilon) , \label{eq:KG_keyleak_4}
\end{align}
for some $\epsilon>0$.
Finally,
\begin{align*}
\bE[\mathsf{L}_k(\cC)] &= \IC{K^{b-1}}{\V{Z}^b}{\cC}\\
&= \HC{K^{b-1}}{\cC} -\HC{K^{b-1}}{\cC \V{Z}^b} \\
&\leq n(b-1)\bar{S}_2 -n(b-1)(\bar{S}_2 -\epsilon) \\
&= n(b-1)\epsilon ,
\end{align*}
and the key is asymptotically secure.

\subsection{Key Uniformity}

The uniformity of the keys is defined in~\eqref{eq:def_sk_unif}. Using~\eqref{eq:KG_keyleak_4}, we obtain
\begin{align*}
 \bE[\mathsf{U}_k(\cC)] &= n(b-1)\bar{S}_2 -\HC{K^{b-1}}{\cC} \\
 &\leq n(b-1)\bar{S}_2 -\HC{K^{b-1}}{\cC \V{Z}^b} \\
 &\leq n(b-1)\epsilon ,
\end{align*}
and thus the key is asymptotically uniform.

\subsection{Information Leakage}

We now proceed to bound the information leakage of the $b-1$ messages
$\bM^b=(\bM_0^b, \bM_1^b)$. Consider first,
\begin{align}
\MoveEqLeft[1]
\IC{\bM_0^b}{\V{Z}^b}{\cC} \nonumber\\
 &= \sum\nolimits_{j=2}^b \IC{\bM_{0j}}{\V{Z}^b}{\cC \bM_0^{j-1}}  \nonumber\\
 &\leq \sum\nolimits_{j=2}^b \IC{\bM_{0j}}{\V{Z}^b \V{T}_j \bM_0^{j-1} L_{1(j-1)} L_{2(j-1)} K_{j-1}'}{\cC} \nonumber\\
 &= \sum\nolimits_{j=2}^b \Big[ \IC{\bM_{0j}}{\V{Z}_j \V{T}_j}{\cC L_{1(j-1)} L_{2(j-1)} K_{j-1}'} \nonumber\\
 &\quad +\IC{\bM_{0j}}{\V{Z}_{j+1}^b}{\cC \V{Z}_j \V{T}_j L_{1(j-1)} L_{2(j-1)} K_{j-1}'} \Big], \label{eq:KG_infleak_1c}
\end{align}
where the last equality is due to $(L_{1(j-1)} L_{2(j-1)} K_{j-1}')$ being 
independent of $\bM_{0j}$ and the Markov chain $(\V{Z}^{j-1} \bM_0^{j-1})
\mkv (\cC L_{1(j-1)} L_{2(j-1)} K_{j-1}') \mkv (\bM_{0j} \V{Z}_j^b)$, see 
Remark~\ref{rk:KG-markov}.

The first term on the right-hand side of~\eqref{eq:KG_infleak_1c} corresponds to the information leakage 
in block $j$ of the message $\bM_{0j}$ given the indices $(L_{j}' L_{j}'')$, 
which is upper-bounded by $n\eta_1$ thanks to~\eqref{eq:KG_rates_6}. 
The conditioning over $K_{j-1}'$ does not affect this term because $\V{Z}_j$ 
is only correlated to $\bM_{1j}'=\bM_{1j}\oplus K_{j-1}'$ which is independent of 
$K_{j-1}'$, given that $\bM_{1j}$ is uniformly distributed on 
$[1:2^{nR_1}]$ and independent of $K_{j-1}'$.

On the other hand, the second term on the right-hand side of~\eqref{eq:KG_infleak_1c} can be bounded as follows
\begin{subequations}\label{eq:KG_infleak_2}
\begin{align}
\MoveEqLeft[1]
\IC{\bM_{0j}}{\V{Z}_{j+1}^b}{\cC \V{Z}_j \V{T}_j L_{1(j-1)} L_{2(j-1)} K_{j-1}'} & \nonumber\\
 &\leq \IC{\bM_{0j} L_{1(j-1)} L_{2(j-1)} K_{j-1}' \V{Z}_j}{\V{Z}_{j+1}^b}{\cC \V{T}_j} \nonumber\\
 &\leq \IC{\V{U}_j \V{Z}_j}{\V{Z}_{j+1}^b}{\cC \V{T}_j} \label{eq:KG_infleak_2a} \displaybreak[2]\\
 &\leq \IC{\V{U}_j \V{Z}_j}{L_{1j} L_{2j} \bM_{1(j+1)}'}{\cC \V{T}_j} \label{eq:KG_infleak_2b}\\
 &= \IC{\V{U}_j \V{Z}_j}{L_{2j}}{\cC \V{T}_j} +\IC{\V{U}_j \V{Z}_j}{\bM_{1(j+1)}'}{\cC \V{T}_j L_{2j}} \nonumber\\
 &\leq \IC{\V{U}_j \V{Z}_j}{L_{2j}}{\cC \V{T}_j} +\I{K_j'}{\bM_{1(j+1)}'} \label{eq:KG_infleak_2c}\\
 &= \IC{\V{U}_j \V{Z}_j}{L_{2j}}{\cC \V{T}_j} , \label{eq:KG_infleak_2d}
\end{align}
\end{subequations}
where
\begin{itemize}
 \item \eqref{eq:KG_infleak_2a} is due to $(\bM_{0j} L_{1(j-1)} L_{2(j-1)} K_{j-1}') \mkv (\cC \V{U}_j) \mkv (\V{T}_j \V{Z}_j^b)$ being a Markov chain since $\V{U}_j$ hides the indices; 
 \item \eqref{eq:KG_infleak_2b} is due to $(\V{U}_j \V{T}_j \V{Z}_j) \mkv (\cC L_{1j} L_{2j} \bM_{1(j+1)}') \mkv \V{Z}_{j+1}^b$, see Remark~\ref{rk:KG-markov};
 \item \eqref{eq:KG_infleak_2c} is due to the Markov chain $\bM_{1(j+1)}' \mkv K_j' \mkv (\cC \V{U}_j \V{T}_j \V{Z}_j L_{2j})$; and,
 \item \eqref{eq:KG_infleak_2d} is again due to $\H{\bM_{1(j+1)}\oplus K_{j}'} = \H{\bM_{1(j+1)}}$.
\end{itemize}

We proceed to bound~\eqref{eq:KG_infleak_2d}, where we remove the block index $j$ for notational simplicity,%
\begin{align}
\MoveEqLeft[1]
 \IC{\V{U} \V{Z}}{L_2}{\cC \V{T}} \nonumber\\
 &= \HC{L_2}{\cC \V{T}} -\HC{L_2}{\cC \V{U} \V{T} \V{Z}} \nonumber\displaybreak[2]\\
 &= \HC{L_2}{\cC \V{T}} -\HC{L_2 K \V{V}}{\cC \V{U} \V{T} \V{Z}} \nonumber\\
 &\quad +\HC{K}{\cC \V{U} \V{T} \V{Z} L_2} +\HC{\V{V}}{\cC \V{U} \V{T} \V{Z} L_2 K} \nonumber\\
 &\leq n\tilde{S}_2 -\HC{\V{V}}{\cC \V{U} \V{T} \V{Z}} +n\bar{S}_2 +n\epsilon_n , \label{eq:KG_infleak_3}
\end{align}
where the inequality follows from bounding the indices $L_2$ and $K$ by their cardinality, and the last entropy by Lemma~\ref{lm:KG-Fano}. The remaining entropy may be bounded using the following lemma.

\vspace{1mm}
\begin{lemma}\label{lm:KG-Bound2}
Let $\eta>0$ and $\epsilon_2$ defined in~\eqref{eq:KG_rates}. Then, given the codebook generation and encoding procedure of the scheme,
\begin{equation}
\HC{\V{V}}{\cC \V{U} \V{T} \V{Z}} \geq n \big[ \IC{V}{X\hat{Y}}{UTZ} +\epsilon_2 -\eta \big].
\end{equation}
for sufficiently large $n$.
\end{lemma}
\begin{IEEEproof}
The proof is found in Appendix~\ref{sec:Proof-Lemma-Bound}.
\end{IEEEproof}
\vspace{1mm}

Using the definitions of $\tilde{S}_2$ and $\bar{S}_2$ from~\eqref{eq:KG_rates}, and Lemma~\ref{lm:KG-Bound2}, we bound~\eqref{eq:KG_infleak_3} as follows
\begin{equation*}
 \IC{\V{U} \V{Z}}{L_2}{\cC \V{T}} \leq n(\tilde{\epsilon}_2 +\epsilon_n +\eta) \triangleq n\eta_2 ,
\end{equation*}
for some $\eta_2>0$,
which let us bound~\eqref{eq:KG_infleak_2}, and in turn, \eqref{eq:KG_infleak_1c},
\begin{equation*}
\IC{\bM_0^b}{\V{Z}^b}{\cC} \leq \sum_{j=2}^b (n\eta_1 +n\eta_2) \triangleq n(b-1)\eta_3 .
\end{equation*}

Now consider,
\begin{align}
\MoveEqLeft[1]
\IC{\bM_1^b}{\V{Z}^b}{\cC \bM_0^b} & \nonumber\\
&= \sum\nolimits_{j=2}^b \IC{\bM_{1j}}{\V{Z}^b}{\cC \bM_0^b \bM_1^{j-1}} \nonumber\\
 &\leq \sum\nolimits_{j=2}^b \IC{\bM_{1j}}{\V{U}_{j-1} \V{T}_{j-1}^j \V{Z}^b}{\cC \bM_0^b \bM_1^{j-1}} \nonumber\\
 &= \sum\nolimits_{j=2}^b \Big[ 
 \IC{\bM_{1j}}{\V{U}_{j-1} \V{T}_{j-1} \V{Z}^{j-1}}{\cC \bM_0^b \bM_1^{j-1}} \nonumber\\
&\quad +\IC{\bM_{1j}}{\V{T}_j \V{Z}_j}{\cC \bM_0^b \bM_1^{j-1} \V{U}_{j-1} \V{T}_{j-1} \V{Z}^{j-1}} \nonumber\\
&\quad +\IC{\bM_{1j}}{\V{Z}_{j+1}^b}{\cC \bM_0^b \bM_1^{j-1} \V{U}_{j-1} \V{T}_{j-1}^j \V{Z}^j} \Big]. \label{eq:KG_infleak_4c}
\end{align}
The first term in~\eqref{eq:KG_infleak_4c} is zero due to the independence
between $(\cC \V{U}_{j-1} \V{T}_{j-1} \V{Z}^{j-1} \bM_0^b \bM_1^{j-1})$ and
$\bM_{1j}$, while the second term can be bounded as follows
\begin{subequations}
\begin{align}
\MoveEqLeft[1]
\IC{\bM_{1j}}{\V{T}_j \V{Z}_j}{\cC \bM_0^b \bM_1^{j-1} \V{U}_{j-1} \V{T}_{j-1} \V{Z}^{j-1}} & \nonumber\\
&\leq \IC{\bM_{1j}}{\bM_{1j}'}{\cC \bM_0^b \bM_1^{j-1} \V{U}_{j-1} \V{T}_{j-1} \V{Z}^{j-1}} \label{eq:KG_infleak_5a} \\
 &\leq \IC{\bM_0^b \bM_1^{j} \V{Z}^{j-2}}{\bM_{1j}'}{\cC \V{U}_{j-1} \V{T}_{j-1} \V{Z}_{j-1}} \nonumber\\
 &= \IC{\bM_0^b \bM_1^{j} \V{Z}^{j-2}}{K_{j-1}'}{\cC \V{U}_{j-1} \V{T}_{j-1} \V{Z}_{j-1}} \nonumber\\
 &\quad +\HC{\bM_{1j}'}{\cC \V{U}_{j-1} \V{T}_{j-1} \V{Z}_{j-1}} \nonumber\\
 &\quad -\HC{K_{j-1}'}{\cC \V{U}_{j-1} \V{T}_{j-1} \V{Z}_{j-1}} \nonumber\\
 &\leq nR_1 -\HC{K_{j-1}}{\cC \V{U}_{j-1} \V{T}_{j-1} \V{Z}_{j-1}} \nonumber\\
 &\quad +\HC{K_{j-1}}{\cC \V{U}_{j-1} \V{T}_{j-1} \V{Z}_{j-1} K_{j-1}'} \label{eq:KG_infleak_5b} \\
%
 &\leq n\bar{S}_2 -\HC{K}{\cC \V{U} \V{T} \V{Z}}, \label{eq:KG_infleak_5c} \\
 &= n\bar{S}_2 -\HC{K L_2 \V{V}}{\cC \V{U} \V{T} \V{Z}} +\HC{L_2}{\cC \V{U} \V{T} \V{Z} K} \nonumber\\
 &\quad  +\HC{\V{V}}{\cC \V{U} \V{T} \V{Z} L_2 K} \nonumber\\
 &\leq n \big[ \bar{S}_2 -\IC{V}{X\hat{Y}}{UTZ} -\epsilon_2 +\eta +\tilde{S}_2 +\epsilon_n \big] \label{eq:KG_infleak_5d} \\
 &= n (\tilde{\epsilon}_2 +\eta +\epsilon_n) ,
\end{align}
\end{subequations}
where
\begin{itemize}
 \item \eqref{eq:KG_infleak_5a} is due to $\bM_{1j} \mkv \bM_{1j}' \mkv (\V{T}_j \V{Z}_j)$ being a Markov chain since $\bM_{1j}' = \bM_{1j}\oplus K_{j-1}'$;
 \item \eqref{eq:KG_infleak_5b} is due to $(\bM_0^b \bM_1^{j} \V{Z}^{j-2}) \mkv (\cC \V{U}_{j-1} \V{T}_{j-1} \V{Z}_{j-1}) \mkv K_{j-1}'$ being a Markov chain and $\H{\bM_{1j}'}=nR_1$;
 \item \eqref{eq:KG_infleak_5c} is due to $\HC{K_{j-1}}{\cC K_{j-1}'}\leq n(\bar{S}_2-R_1)$ and the block index $j$ being removed for brevity; and,
 \item \eqref{eq:KG_infleak_5d} follows similar steps as~\eqref{eq:KG_infleak_3}.
\end{itemize}

The third term in~\eqref{eq:KG_infleak_4c} may be bounded from above as follows
\begin{align*}
\MoveEqLeft[1]
\IC{\bM_{1j}}{\V{Z}_{j+1}^b}{\cC \bM_0^b \bM_1^{j-1} \V{U}_{j-1} \V{T}_{j-1}^j \V{Z}^j} & \nonumber\\
&\leq \IC{\bM_{1j}}{L_{1j} L_{2j} \bM_{1(j+1)}'}{\cC \bM_0^b \bM_1^{j-1} \V{U}_{j-1} \V{T}_{j-1}^j \V{Z}^j} \\
 &\leq \IC{\bM_0^b \bM_1^{j} \V{U}_{j-1} \V{T}_{j-1} \V{Z}^j}{L_{1j} L_{2j} \bM_{1(j+1)}'}{\cC \V{T}_j} \\
 &\leq \IC{\V{U}_j \V{Z}_j}{L_{1j} L_{2j} \bM_{1(j+1)}'}{\cC \V{T}_j} \\
 &\leq n\eta_2 ,
\end{align*}
where the last inequality is bounded exactly as~\eqref{eq:KG_infleak_2b}. Thus, 
\eqref{eq:KG_infleak_4c} is upper-bounded as
\begin{equation*}
\IC{\bM_1^b}{\V{Z}^b}{\cC \bM_0^b} \leq \sum_{j=2}^b 2n\eta_2 = 2n(b-1)\eta_2 .
\end{equation*}

Finally, the total information leakage is
\begin{align*}
\bE[\mathsf{L}(\cC)] &= \IC{\bM_0^b \bM_1^b}{\V{Z}^b}{\cC} \\
&= \IC{\bM_0^b}{\V{Z}^b}{\cC} +\IC{\bM_1^b}{\V{Z}^b}{\cC \bM_0^b} \\
&\leq n(b-1)(2\eta_2 +\eta_3) ,
\end{align*}
which assures that the eavesdropper has negligible knowledge of the messages asymptotically.

\subsection{Sufficient Conditions (\texorpdfstring{$R_{KG_1}$}{RKG1})}

Putting all the pieces together, we have proved that the proposed scheme allows the
encoder to transmit a message uniformly distributed in $[1:2^{nR}]$, $R =
R_{KG_1} = R_0 +R_1$, while keeping it secret from the eavesdropper if
\begin{align*}
 \IC{V}{Z}{UT} &\leq \IC{V}{Y}{UT} , \\
 \tilde{S}' +\tilde{S}'' = \tilde{S}_1 +\tilde{S}_2 &= \IC{V}{X\hat{Y}}{UY} +\epsilon_{12} , \\
 R_1 \leq \bar{S}_2 &= \IC{V}{Y}{UT} - \IC{V}{Z}{UT} , \\
 \tilde{S}'' +R_0 +R_1 +R_f &< \IC{U}{Y}{Q} -\delta , \\
 \tilde{S}' +\tilde{S}'' +R_0 +R_1 +R_f &< \I{U}{Y} -\delta , \\
 R_1 +R_f &= \IC{U}{Z}{Q} +\IC{U}{T}{QZ} -\epsilon' ,
\end{align*}
where $\epsilon_{12} =\epsilon_1 +\epsilon_2 +\tilde{\epsilon}_1 +\tilde{\epsilon}_2$.
After applying Fourier-Motzkin elimination, we obtain the bounds in~\eqref{eq:KG_region1_rate} subject to
\begin{subequations}\label{eq:KG_conds}
\begin{align}
 \IC{V}{Z}{UT} &\leq \IC{V}{Y}{UT} , \label{eq:KG_conds_1} \\
 \IC{U}{TZ}{Q} &\leq \IC{U}{Y}{Q} , \label{eq:KG_conds_2} \\
 \IC{V}{X\hat{Y}}{UY} +\IC{U}{TZ}{Q} &\leq \I{U}{Y} . \label{eq:KG_conds_3}
\end{align}
\end{subequations}
Nonetheless, these conditions are redundant after the maximization process. If
for a certain PD, condition~\eqref{eq:KG_conds_1} is not satisfied, then,
$R_{KG_1}$ with $T=V=\emptyset$ attains a higher value. Similarly, if
either~\eqref{eq:KG_conds_2} or~\eqref{eq:KG_conds_3} does not hold for a
certain PD, then, $R_{KG_2}$ with $Q=\emptyset$ attains a higher value.

We have shown thus far that, \emph{averaged} over all possible codebooks, the probability of error, the key leakage and (non-)uniformity, and the information leakage rate become negligible as $(n,b)\to\infty$ if conditions~\eqref{eq:KG_region1_rate} hold true. Nonetheless, by applying the selection lemma~\cite[Lemma 2.2]{bloch_physical_2011}, we may conclude that there exists a \emph{specific} sequence of codebooks such that the probability of error, the key leakage and (non-)uniformity, and the information leakage rate tend to zero as $(n,b)\to\infty$.

The bounds on the cardinality of the alphabets $\cQ$, $\cU$, $\cT$, and $\cV$ follow from Fenchel--Eggleston--Carath{\'e}odory's theorem and the standard cardinality bounding technique~\cite[Appendix C]{gamal_network_2011}; therefore their proof is omitted.

\subsection{Achievable Rate \texorpdfstring{$R_{KG_2}$}{RKG2}}\label{ssec:KG_KG2}

The second strategy tackles the situation where the eavesdropper experiences a
better channel than the legitimate receiver and can therefore decode
everything sent by the encoder. In $R_{KG_1}$, when either the
condition~\eqref{eq:KG_conds_2} or~\eqref{eq:KG_conds_3} is not satisfied, the
rate of the unencrypted message ($R_0$) is negative. Therefore, in this second
strategy the message is encrypted completely. The proof is similar to the one of
$R_{KG_1}$ and we only point out the differences in what follows.

\subsubsection{Codebook Generation}

Since the eavesdropper is able to decode everything, there is no need for the
codeword $\V{q}(\cdot)$ as a lower layer for $\V{u}(\cdot)$, which in turn makes 
the bit recombination $(l_j', l_j'') = M_l(l_{1(j-1)},l_{2(j-1)})$ unnecessary.
Additionally, since the encoder cannot send the message without encrypting it,
$R_0=0$ and $R_f=0$, and the condition~\eqref{eq:KG_rates_6} disappears.
We therefore take the joint distribution~\eqref{eq:KG_pmf} with $Q=\emptyset$
and build the codebooks for each block as in Appendix~\ref{ssec:KG_code} without
$\V{q}(\cdot)$ and with $\V{t}(\cdot)$ superimposed over $\V{u}(\cdot)$. The
quantities~\eqref{eq:KG_rates} are modified as follows:
\begin{align*}
S_1         &= \IC{T}{X\hat{Y}}{U} +\epsilon_1 , \\
\tilde{S}_1 &= \IC{T}{X\hat{Y}}{U} -\IC{T}{Y}{U} +\epsilon_1 +\tilde{\epsilon}_1 .
\end{align*}
%

\subsubsection{Encoding and Decoding}

These steps are analogous to the previous proof with two main differences.
First, there is no bit recombination in the transmission of the bin indices. 
Second, the encoder only sends an encrypted message $m_j'=m_j\oplus k_{j-1}'$ 
using the key obtained from the feedback of the previous block. Briefly, if 
$\V{t}(\ve{r}_{j-1},s_{1(j-1)})\in B_1(l_{1(j-1)})$ and $\V{v}(\ve{r}_{j-1}, 
s_{1(j-1)},s_{2(j-1)}) \in \bar{B}_2(s_{1(j-1)},l_{2(j-1)},k_{j-1})$, the 
encoder sends $\V{u}(l_{1(j-1)},l_{2(j-1)}, m_j') = \V{u}(\ve{r}_j)$ during block $j$.

\subsubsection{Key and Information Leakage}

The proof for the key secrecy and uniformity is the same while the one for the information
leakage is simplified. Since there is no unencrypted message, i.e., $R_0=0$ the
bounding of $\IC{\bM_0^b}{\V{Z}^b}{\cC}$ becomes trivial and the
condition~\eqref{eq:KG_rates_6} is no longer necessary.

\subsubsection{Final Expression}

The sufficient conditions in this second strategy for the encoder to transmit a
message uniformly distributed in $[1:2^{nR}]$, $R=R_{KG_2}$, while keeping it secret 
from the eavesdropper are given by
\begin{subequations}\label{eq:KG_kg2_bounds}
\begin{align}
 \IC{V}{Z}{UT} &\leq \IC{V}{Y}{UT} , \\
 \tilde{S}_1 +\tilde{S}_2 &= \IC{V}{X\hat{Y}}{UY} +\epsilon_1 +\epsilon_2 +\tilde{\epsilon}_1 +\tilde{\epsilon}_2 ,  \displaybreak[2]\\
 R \leq \bar{S}_2 &= \IC{V}{Y}{UT} - \IC{V}{Z}{UT} , \\
 \tilde{S}_1 +\tilde{S}_2 +R &< \I{U}{Y} -\delta ,
\end{align}
\end{subequations}
which yields~\eqref{eq:KG_region2_rate} after applying Fourier Motzkin elimination.%

\subsection{Final Remarks}\label{ssec:KG_Final}

The preceding proof guarantees that there exists a specific $(2^{nR},n)$ code $\mathsf{c}_n$ whose rate is achievable under the \emph{weak secrecy} condition~\eqref{eq:s_rate_weak}.
Nevertheless, using the method proposed in~\cite{maurer_strong_2000}, we can show that the achievable secrecy rate also complies with the \emph{strong secrecy} condition~\eqref{eq:s_rate_strong}.
In the sequel, we show how this is achieved following~\cite[Prop.~4.10]{bloch_physical_2011}.

Let $\epsilon>0$ and consider a code $\mathsf{c}_n$ with rate 
\begin{equation}
R = \max\left\{ \max_{p\in\cP_{I_1}} R_{KG_1}(p), \ \max_{p'\in\cP_{I_2}} R_{KG_2}(p') \right\} -\epsilon ,  \label{eq:KG_strong-1}
\end{equation}
where the definitions for the rates are found in~\eqref{eq:KG_region1_rate} and~\eqref{eq:KG_region2_rate},
such that condition~\eqref{eq:s_rate_weak} holds.
The encoder then uses this code $m$ times\footnote{The proof of the scheme is based on splitting the transmission in $b$ blocks of $n$ channel uses; thus, the whole weakly secret transmission takes place in $nb$ channel uses. To simplify the presentation of this part, we consider that a weakly secret transmission, i.e., each of the $m$ times the code $\mathsf{c}_n$ is employed, takes place in $n$ channel uses.} %
to transmit $m$ independent messages. In each transmission $i\in[1:m]$, the encoder transmits $\bM_i$, the decoder obtains $\hat{\bM}_i$, and the eavesdropper observes $\V{Z}_i$. This situation is akin to the ``source model'' in the problem of secret key generation where the encoder, the decoder, and the eavesdropper observe $m$ realizations of the random variables
\begin{align}
 X'&\triangleq \bM , &
 Y'&\triangleq \hat{\bM} , &
 &\textnormal{and} &
 Z'&\triangleq \V{Z} , \label{eq:KG_strong-2}
\end{align}
respectively.
According to~\cite[Thm.~4.7]{bloch_physical_2011} and for some $\epsilon'>0$, the legitimate users can agree on a strong secret key $\bar{K}$ of length
\begin{equation*}
 k = m \big[ \I{X'}{Y'} -\I{X'}{Z'} -\epsilon' \big] \geq mn (R -\epsilon'') , 
\end{equation*}
where the inequality follows, for some $\epsilon''>0$, from the definitions in~\eqref{eq:KG_strong-2}, the condition~\eqref{eq:s_rate_weak}, and the fact that the rate of $\bM$ is determined by~\eqref{eq:KG_strong-1}.

The strong secret key $\bar{K}$ is obtained by means of a one-way direct reconciliation protocol and privacy amplification with extractors. These two steps involve the transmission of additional information through the channel; in particular, the one-way reconciliation protocol needs $m[\HC{X'}{Y'}+\delta]$ bits of communication and the privacy amplification, $m\delta'$ bits, for some $\delta,\delta'>0$.
Nonetheless, these additional $m'$ channel uses are negligible compared to the total transmission time for large $m$ and $n$, i.e., $m'\leq mn\delta''$, for some small $\delta''> 0$; thus, the rate of the strong secret key $\bar{K}$ is bounded from below as
\begin{equation*}
 \frac{k}{mn+m'} \geq R -\bar\epsilon , 
\end{equation*}
for some $\bar\epsilon>0$. We refer the reader to~\cite[Sec.~4.5]{bloch_physical_2011} for the details.

Lastly, it remains to be seen if the secret key $\bar{K}$ can be interpreted as a message. Given that all the transmissions are one-way, it is possible for the encoder to choose the key $\bar{K}$ ahead of time and ``invert'' the reconciliation and privacy amplification processes; the encoder then obtains the $m$ messages to transmit using the weak code $\mathsf{c}_n$. Therefore, the final strong secret-key $\bar{K}$ can be treated as a message $\bM$ that satisfies the strong secrecy condition~\eqref{eq:s_rate_strong}.
This concludes the proof of Theorem~\ref{theo-KG}.
\hfill\IEEEQED

\section{Proof of Theorem~\ref{cor-key} (SK Rate Lower Bound)}
\label{sec:Proof_Cor-KG}

In this scheme, the encoder is not interested in transmitting a message but rather agreeing on a secret key with the legitimate receiver. As in the proof of Theorem~\ref{theo-KG}, the encoder splits the transmission in $b$ blocks of $n$ channel uses and employs one of two available strategies to generate the shared secret key.

In the first strategy, the secret key has two components: one is sent over the channel and is kept secret from the eavesdropper by using Wyner's wiretap coding scheme, while the second component is generated thanks to the correlation between the outputs $Y$ and $\hat{Y}$.
On the other hand, the second strategy generates a secret key only relying on the correlation between the channel outputs.

In the following, we present a brief sketch of the proof for both strategies given the similarities with respect to the proof of Theorem~\ref{theo-KG} in Appendix~\ref{sec:Proof_KG}. Consequently, the secret key rate is achievable according to the weak secrecy condition~\eqref{eq:sk_rate_weak} but we show at the end of this Appendix that the strong secrecy condition~\eqref{eq:sk_rate_strong} also holds true.

\subsection{First Strategy}

This part follows the same steps as the proof of the achievable secrecy rate $R_{KG_1}$, found in Appendix~\ref{sec:Proof_KG}, but without the transmission of an encrypted message.
Thus, at the end of the $b$ transmission blocks, the encoder and the legitimate receiver will agree with high probability on a key of rate $\frac{b-1}{b} R_k$.
Due to the similarity with the proof of $R_{KG_1}$, we only point out the differences in the sequel.

\subsubsection{Codebook Generation}

The codebook is generated in the same way as for the achievable rate $R_{KG_1}$, with 
the exception of the codeword $\V{u}(\cdot)$. Specifically, the message $m_0$ 
carried by that scheme becomes a part of the secret key here, i.e., $R_0 = 
R_{k0}$, and the key generated through the feedback link is not used to encrypt
a message but rather becomes the second part of the secret key, i.e., $R_1 = 0$,
$R_{k1}=\bar{S}_2$, and $R_f = \IC{U}{TZ}{Q} -\epsilon'$ replaces
\eqref{eq:KG_rates_6}.

Step~\ref{it:kg-u} in Appendix~\ref{ssec:KG_code} thus becomes:
\begin{enumerate}
 \item[2)] For each $\V{q}(l')$, randomly pick $2^{n(\tilde{S}''+R_{k0}+R_f)}$ sequences $\V{u}(\ve{r}) \equiv \V{u}(l',l'',k_0,l_f)$, where $l'' \in [1:2^{n\tilde{S}''}]$, $k_0 \in [1:2^{nR_{k0}}]$, and $l_f \in [1:2^{nR_f}]$, from $\typ{U|\V{q}(l')}$.
\end{enumerate}

\subsubsection{Encoding and Decoding}

These steps are similar to the ones for the achievable rate $R_{KG_1}$ but no message is transmitted.
In each block $j \in [2:b]$, the encoder chooses uniformly at random a key index $k_{0j} \in [1:2^{nR_{k0}}]$ and a noise index $l_{fj} \in [1:2^{nR_f}]$.
It then sends these indices, along with the bin indices $(l_j', l_j'')$ of the description of the previous block's feedback sequence, through the codeword $\V{u}(l_j', l_j'', k_{0j}, l_{fj}) = \V{u}(\ve{r}_j)$.

\subsubsection{Key and Information Leakage}

The proof for the key leakage of the achievable rate $R_{KG_1}$ assures that the part of the key that is created using the description, i.e., $k_1$, is kept secret from the eavesdropper, while the proof of the information leakage guarantees that the part that is sent through the codeword $\V{u}(\ve{r})$, i.e., $k_0$, is also secure. 
Both proofs get simplified since $k_1$ is not used to encrypt a message, and, therefore, it is not transmitted. Remark~\ref{rk:KG-markov} should now state that only the variables $L_{1j}$ and $L_{2j}$ are responsible for the correlation between blocks.

\subsubsection{Key Uniformity}

The encoding procedure states that the first part of the key, i.e., $k_0$, is chosen uniformly at random, while the proof of the key uniformity of the achievable rate $R_{KG_1}$ assures that the other part, i.e., $k_1$, is asymptotically uniform.

\subsubsection{Final Expression}

The sufficient conditions in this first strategy, which allows the legitimate users to agree upon a key uniformly distributed in $[1:2^{nR_k}]$, $R_k = R_{k0} +R_{k1}$, while keeping it secret from the eavesdropper, are
\begin{align*}
 \IC{V}{Z}{UT} &\leq \IC{V}{Y}{UT} , \\
 \tilde{S}' +\tilde{S}'' = \tilde{S}_1 +\tilde{S}_2 &= \IC{V}{X\hat{Y}}{UY} +\epsilon_1 +\epsilon_2 +\tilde{\epsilon}_1 +\tilde{\epsilon}_2 , \\
 R_{k1} \leq \bar{S}_2 &= \IC{V}{Y}{UT} -\IC{V}{Z}{UT} , \displaybreak[2]\\
 \tilde{S}'' +R_{k0} +R_f &< \IC{U}{Y}{Q} -\delta , \\
 \tilde{S}' +\tilde{S}'' +R_{k0} +R_f &< \I{U}{Y} -\delta , \\
 R_f &= \IC{U}{Z}{Q} +\IC{U}{T}{QZ} -\epsilon' .
\end{align*}
After applying Fourier Motzkin elimination to this set of inequalities, we obtain
\begin{multline}
R_k\leq  \I{U}{Y} -\IC{U}{Z}{Q} +\IC{V}{Y}{UT} -\IC{V}{Z}{UT}\quad\ \\
-\IC{U}{T}{QZ} -\max\{ \I{Q}{Y}, \, \IC{V}{X\hat{Y}}{UY} \} , \label{eq:Cor-KG_kg1_rate}
\end{multline}
subject to the conditions~\eqref{eq:KG_conds}. However, these conditions are
redundant after the maximization process as in $R_{KG_1}$.

\subsection{Second Strategy}

This part is derived from the achievable rate $R_{KG_2}$, where we are only interested in generating a secret key, i.e., $R_k\leq\bar{S}_2$. As before, the encoder does not transmit an encrypted message, i.e., $R=0$, and the codeword $\V{u}(\cdot)$ is modified accordingly. Refer to Appendix~\ref{ssec:KG_KG2} for details.

The sufficient conditions in this second strategy are derived from~\eqref{eq:KG_kg2_bounds}:%
\begin{align*}
 \IC{V}{Z}{UT} &\leq \IC{V}{Y}{UT} , \\
 \tilde{S}_1 +\tilde{S}_2 &= \IC{V}{X\hat{Y}}{UY} +\epsilon_1 +\epsilon_2 +\tilde{\epsilon}_1 +\tilde{\epsilon}_2 , \\
 R_k \leq \bar{S}_2 &= \IC{V}{Y}{UT} - \IC{V}{Z}{UT} , \\
 \tilde{S}_1 +\tilde{S}_2 &< \I{U}{Y} -\delta .
\end{align*}
After applying Fourier Motzkin elimination to this system, we obtain
\begin{equation}
R_k \leq \IC{V}{Y}{UT} -\IC{V}{Z}{UT} \label{eq:Cor-KG_kg2_rate}
\end{equation}
subject to the condition
\begin{equation}
\IC{V}{X\hat{Y}}{UY} \leq \I{U}{Y} . \label{eq:Cor-KG_kg2_cond}
\end{equation}

\subsection{Final Remarks}

The final achievable secret key rate $R_k$, which is the union of~\eqref{eq:Cor-KG_kg1_rate} and~\eqref{eq:Cor-KG_kg2_rate} conditioned on~\eqref{eq:Cor-KG_kg2_cond} and maximized over all possible joint PDs, can be succinctly written
as~\eqref{eq:cor-key_rate} and~\eqref{eq:cor-key_cond}.
As in the proof of Theorem~\ref{theo-KG}, the preceding rate was shown to be achievable under the \emph{weak secrecy} condition~\eqref{eq:sk_rate_weak}. Nonetheless, following the same procedure as in Appendix~\ref{ssec:KG_Final}, we can show that said rate is also achievable under the \emph{strong secrecy} condition~\eqref{eq:sk_rate_strong}.

In short, the encoder employs the previously described SK code $\mathsf{c}_n$ $m$ times and the legitimate users agree on $m$ weakly secure keys.
These keys may be considered as $m$ observations of correlated sources and, similarly to~\cite[Prop.~4.10]{bloch_physical_2011}, they may be further distilled to obtain a strong secret key by means of information reconciliation and privacy amplification with extractors.
The proof is a simplified version of the one presented in Appendix~\ref{ssec:KG_Final}, and thus we omit it here.
The main difference is the absence of a transmitted message, which eliminates the need to ``invert'' the reconciliation and privacy amplification processes.
This concludes the proof of Theorem~\ref{cor-key}.
\hfill\IEEEQED

\section{Proof of Theorem~\ref{theo-OB} (Secrecy Rate Upper Bound)}
\label{sec:Proof-OB}

Let $R$ be an achievable strong secrecy rate according to Definition~\ref{def:s_rate} with the appropriate modifications for the model with parallel sources.
Then, for $\epsilon>0$ and sufficiently large $n$, 
there exist functions enc$_i(\cdot)$ and dec$(\cdot)$ such that
\begin{subequations}\label{eq:OB_functions}
\begin{align}
X_{ci}      &= \textnormal{enc}_i(\bM_n, R_r, \hat{Y}_s^{i-1} ) , \label{eq:OB_functions1}\\
\hat{\bM}_n &= \textnormal{dec}( Y_s^n, Y_c^n ) ,
\end{align}
\end{subequations}
which verify
\begin{align}
\PR{ \smash{\hat{\bM}_n} \neq \bM_n } &\leq \epsilon , \label{eq:OB_cond1} \\
\I{\bM_n}{Z_s^n Z_c^n} &\leq \epsilon , \label{eq:OB_cond2}
\end{align}
where we have dropped the conditioning on the codebook $\mathsf{c}_n$ from~\eqref{eq:OB_cond2} and all subsequent calculations for clarity.

First consider,
\begin{subequations}\label{eq:OB_rate1}
\begin{align}
nR &= \H{\bM_n} \nonumber\\
 &= \HC{\bM_n}{Z_s^n Y_c^n} +\I{\bM_n}{Z_s^n Y_c^n} \nonumber\\
 &\leq \HC{\bM_n}{Z_s^n Y_c^n} +\I{\bM_n}{Z_s^n Y_c^n} -\I{\bM_n}{Z_s^n Z_c^n} +\epsilon \label{eq:OB_rate1a} \\
 &= \HC{\bM_n}{Z_s^n Y_c^n} +\IC{\bM_n}{Y_c^n}{Z_s^n} -\IC{\bM_n}{Z_c^n}{Z_s^n} +\epsilon \nonumber\displaybreak[2]\\
 &\leq \HC{\bM_n}{Z_s^n Y_c^n} -\HC{\bM_n}{Y_s^n Y_c^n} \nonumber\\
&\quad +\IC{\bM_n}{Y_c^n}{Z_s^n} -\IC{\bM_n}{Z_c^n}{Z_s^n} +n\epsilon_n \label{eq:OB_rate1b} \displaybreak[2]\\
%
 &= {\underbrace{\IC{\bM_n}{Y_s^n}{Y_c^n} -\IC{\bM_n}{Z_s^n}{Y_c^n}}_{R_s}}\nonumber\\
&\quad +{\underbrace{\IC{\bM_n}{Y_c^n}{Z_s^n} -\IC{\bM_n}{Z_c^n}{Z_s^n}}_{R_c}} +n\epsilon_n , \label{eq:OB_rate1c}
\end{align}
\end{subequations}
where
\begin{itemize}
 \item \eqref{eq:OB_rate1a} is due to the security condition~\eqref{eq:OB_cond2}; and,
 \item \eqref{eq:OB_rate1b} follows from~\eqref{eq:OB_functions}, \eqref{eq:OB_cond1}, and Fano's inequality, $\HC{\bM_n}{Y_s^n Y_c^n} \leq n\epsilon_n'$.
\end{itemize}
We now study separately the ``source'' term $R_s$ and the ``channel'' term $R_c$.
\begin{subequations}\label{eq:OB_rate2}
\begin{align}
R_s &= \sum\nolimits_{i=1}^n \IC{\bM_n}{Y_{si}}{Y_c^n Y_s^{i-1}} -\IC{\bM_n}{Z_{si}}{Y_c^n Z_{s(i+1)}^n} \displaybreak[2]\nonumber\\
 &= \sum\nolimits_{i=1}^n \IC{\bM_n}{Y_{si}}{Y_c^n Y_s^{i-1} Z_{s(i+1)}^n} \nonumber\\
&\qquad -\IC{\bM_n}{Z_{si}}{Y_c^n Y_s^{i-1} Z_{s(i+1)}^n} \displaybreak[2]\label{eq:OB_rate2a} \\
 &= \sum\nolimits_{i=1}^n \IC{V_i}{Y_{si}}{T_i} -\IC{V_i}{Z_{si}}{T_i} \displaybreak[2]\label{eq:OB_rate2b} \\
 &= n \big[ \IC{V_J}{Y_{sJ}}{T_J J} -\IC{V_J}{Z_{sJ}}{T_J J} \big] \label{eq:OB_rate2c} \\
 &= n \big[ \IC{V}{Y_s}{T} -\IC{V}{Z_s}{T} \big] , \label{eq:OB_rate2d}
\end{align}
\end{subequations}
where
\begin{itemize}
 \item \eqref{eq:OB_rate2a} is due to Csisz\'ar sum identity; 
 \item \eqref{eq:OB_rate2b} stems from the definition of the auxiliary RVs $T_i = (Y_c^n Y_s^{i-1} Z_{s(i+1)}^n)$ and $V_i = (\bM_n T_i)$;
 \item in~\eqref{eq:OB_rate2c} we add the auxiliary RV $J$ uniformly distributed on $[1:n]$ and independent of all the other variables; and,
 \item \eqref{eq:OB_rate2d} follows from the definition of random variables $T = (T_J J)$, $V = (V_J J)$, $Y_s = Y_{sJ}$, and $Z_s = Z_{sJ}$.
\end{itemize}
This establishes the ``source'' term in~\eqref{eq:OB_rate1c} with auxiliary RVs $(TV)$ that satisfy the following Markov chain
\begin{equation}
 T_i \mkv V_i \mkv \hat{Y}_{si} \mkv (Y_{si} Z_{si}) . \label{eq:OB_mkv1}
\end{equation}
The first part of~\eqref{eq:OB_mkv1} is trivial given the definition $V_i =
(\bM_n T_i)$, whereas the second part follows from the i.i.d. nature of the
sources and that they are correlated to the main channel only through the
encoder's input~\eqref{eq:OB_functions1},
\begin{equation*}
(\bM_n Y_c^n Y_s^{i-1} Z_{s(i+1)}^n) \mkv \hat{Y}_{si} \mkv (Y_{si} Z_{si}) . 
\end{equation*}

The ``channel'' term $R_c$ can be single-letterized similarly,
\begin{subequations}\label{eq:OB_rate3}
\begin{align}
R_c &= \sum\nolimits_{i=1}^n \IC{\bM_n}{Y_{ci}}{Z_s^n Y_c^{i-1}} -\IC{\bM_n}{Z_{ci}}{Z_s^n Z_{c(i+1)}^n} \nonumber\\
 &= \sum\nolimits_{i=1}^n \IC{\bM_n}{Y_{ci}}{Z_s^n Y_c^{i-1} Z_{c(i+1)}^n} \nonumber\\
&\qquad-\IC{\bM_n}{Z_{ci}}{Z_s^n Y_c^{i-1} Z_{c(i+1)}^n} \label{eq:OB_rate3a} \\
 &= \sum\nolimits_{i=1}^n \IC{U_i}{Y_{ci}}{Q_i} -\IC{U_i}{Z_{ci}}{Q_i} \label{eq:OB_rate3b} \displaybreak[2]\\
 &= n \big[ \IC{U_L}{Y_{cL}}{Q_L L} -\IC{U_L}{Z_{cL}}{Q_L L} \big] \label{eq:OB_rate3c} \\
 &= n \big[ \IC{U}{Y_c}{Q} -\IC{U}{Z_c}{Q} \big] , \label{eq:OB_rate3d}
\end{align}
\end{subequations}
where
\begin{itemize}
 \item \eqref{eq:OB_rate3a} is due to Csisz\'ar sum identity; 
 \item \eqref{eq:OB_rate3b} stems from the definition of the auxiliary RVs $Q_i = (Z_s^n Y_c^{i-1} Z_{c(i+1)}^n)$ and $U_i = (\bM_n Q_i)$;
 \item in~\eqref{eq:OB_rate3c} we add the auxiliary RV $L$ uniformly distributed on $[1:n]$ and independent of all the other variables; and,
 \item \eqref{eq:OB_rate3d} follows from the definition of random variables $Q = (Q_L L)$, $U = (U_L L)$, $Y_c = Y_{cL}$, and $Z_c = Z_{cL}$. 
\end{itemize}
The auxiliary RVs in this term, i.e., $(QU)$, satisfy the following Markov chain
\begin{equation*}
 Q_i \mkv U_i \mkv X_{ci} \mkv (Y_{ci} Z_{ci}) , 
\end{equation*}
where the nontrivial part is due to the memorylessness property of the channel and~\eqref{eq:OB_functions1}.
Since neither $Q$ nor $U$ appear on other parts of the upper bound, we may expand $R_c$ as
\begin{align}
 R_c &= n \sum_{q\in\cQ} p_Q(q) \big[ \IC{U}{Y_c}{Q=q} -\IC{U}{Z_c}{Q=q} \big] \displaybreak[2]\nonumber\\
 &\leq n \max_{q\in\cQ} \big[ \IC{U}{Y_c}{Q=q} -\IC{U}{Z_c}{Q=q} \big] \nonumber\\
 &= n \big[ \I{U^\star}{Y_c} -\I{U^\star}{Z_c} \big] , \label{eq:OB_rate3bis}
\end{align}
where in the last step we set the auxiliary RV $U^\star \sim p_{U\vert Q}(\cdot\vert q)$ with the specific $q$ that maximizes the preceding expression.

Putting~\eqref{eq:OB_rate1}, \eqref{eq:OB_rate2}, and~\eqref{eq:OB_rate3bis} together, letting $n\to\infty$, and taking arbitrarily small $\epsilon_n$, we obtain the bound~\eqref{eq:theo_OB-rate1}.

In order to obtain~\eqref{eq:theo_OB-rate2}, consider the following,
\begin{subequations}\label{eq:OB_rate4}
\begin{align}
\MoveEqLeft[1]
n(R -\epsilon_n) \nonumber\\
&\leq \I{\bM_n}{Y_s^n Y_c^n} \label{eq:OB_rate4a} \\
&= \I{\bM_n}{\hat{Y}_s^n Y_s^n Y_c^n} -\IC{\bM_n}{\hat{Y}_s^n}{Y_s^n Y_c^n} \nonumber\\
&= \IC{\bM_n}{Y_c^n}{\hat{Y}_s^n} -\IC{\bM_n}{\hat{Y}_s^n}{Y_s^n Y_c^n} \label{eq:OB_rate4b} \\
&= \I{\bM_n \hat{Y}_s^n}{Y_c^n} -\I{\hat{Y}_s^n}{Y_c^n} -\IC{\bM_n}{\hat{Y}_s^n}{Y_s^n Y_c^n} \nonumber\\
&\leq \I{\bM_n \hat{Y}_s^n}{Y_c^n} -\IC{\hat{Y}_s^n}{Y_c^n}{Y_s^n}
-\IC{\bM_n}{\hat{Y}_s^n}{Y_s^n Y_c^n} \label{eq:OB_rate4c} \\
&= \I{\bM_n \hat{Y}_s^n}{Y_c^n} -\IC{\bM_n Y_c^n}{\hat{Y}_s^n}{Y_s^n} \nonumber\\
&\leq \I{X_c^n}{Y_c^n} -\IC{\bM_n Y_c^n}{\hat{Y}_s^n}{Y_s^n} \label{eq:OB_rate4d} \\
&\leq n \I{X_c}{Y_c} -\IC{\bM_n Y_c^n}{\hat{Y}_s^n}{Y_s^n} , \label{eq:OB_rate4e}
\end{align}
\end{subequations}
where
\begin{itemize}
 \item \eqref{eq:OB_rate4a} stems from Fano's inequality;
 \item \eqref{eq:OB_rate4b} and~\eqref{eq:OB_rate4c} follow from $\hat{Y}_s^n$ being independent of $\bM_n$ and the Markov chain $Y_s^n \mkv \hat{Y}_s^n \mkv (\bM_n Y_c^n)$;
 \item \eqref{eq:OB_rate4d} stems from the encoding procedure~\eqref{eq:OB_functions1}; and, 
 \item \eqref{eq:OB_rate4e} is due to the channel being memoryless.
\end{itemize}
The second term in~\eqref{eq:OB_rate4e} can be lower-bounded as follows,
\begin{subequations}\label{eq:OB_rate5}
\begin{align}
\MoveEqLeft[1]
\IC{\bM_n Y_c^n}{\hat{Y}_s^n}{Y_s^n} \nonumber\\
&= \IC{\bM_n Y_c^n}{\hat{Y}_s^n Z_s^n}{Y_s^n} \label{eq:OB_rate5a} \\
 &= \sum\nolimits_{i=1}^n \IC{\bM_n Y_c^n}{\hat{Y}_{si} Z_{si}}{Y_s^n \hat{Y}_{s(i+1)}^n Z_{s(i+1)}^n} \nonumber\\
 &\geq \sum\nolimits_{i=1}^n \IC{\bM_n Y_c^n Y_s^{i-1} Z_{s(i+1)}^n}{\hat{Y}_{si} Z_{si}}{Y_{si}} \label{eq:OB_rate5b} \displaybreak[2]\\
 &= \sum\nolimits_{i=1}^n \IC{V_i}{\hat{Y}_{si} Z_{si}}{Y_{si}} \label{eq:OB_rate5c} \displaybreak[2]\\
 &\geq \sum\nolimits_{i=1}^n \IC{V_i}{\hat{Y}_{si}}{Y_{si}} \displaybreak[2] \nonumber\\
 &= n \IC{V_J}{\hat{Y}_{sJ}}{Y_{sJ} J} \label{eq:OB_rate5d} \displaybreak[2]\\
 &= n \IC{V_J J}{\hat{Y}_{sJ}}{Y_{sJ}} \label{eq:OB_rate5e} \\
 &= n \IC{V}{\hat{Y}_s}{Y_s} , \label{eq:OB_rate5f}
\end{align}
\end{subequations}
where
\begin{itemize}
 \item \eqref{eq:OB_rate5a} is due to $Z_s^n \mkv (Y_s^n \hat{Y}_s^n) \mkv (\bM_n Y_c^n)$; 
 \item \eqref{eq:OB_rate5b} follows from the sources being i.i.d., i.e., $(\hat{Y}_{si} Z_{si}) \mkv Y_{si} \mkv (Y_s^{i-1} Y_{s(i+1)}^n \hat{Y}_{s(i+1)}^n Z_{s(i+1)}^n)$;
 \item in~\eqref{eq:OB_rate5c} we introduce the auxiliary RV $V_i$, see~\eqref{eq:OB_rate2b}; 
 \item in~\eqref{eq:OB_rate5d} we introduce the auxiliary RV $J$, see~\eqref{eq:OB_rate2c}; 
 \item \eqref{eq:OB_rate5e} is due to the independence of $J$ and $(\hat{Y}_{sJ} Y_{sJ})$; and,
 \item \eqref{eq:OB_rate5f} stems from the definition of random variables $V = (V_J J)$, $Y_s = Y_{sJ}$, and $\hat{Y}_s = \hat{Y}_{sJ}$.
\end{itemize}

Putting~\eqref{eq:OB_rate4} and~\eqref{eq:OB_rate5} together, letting $n\rightarrow\infty$, and taking an arbitrarily small $\epsilon_n$, we obtain the bound~\eqref{eq:theo_OB-rate2}.

Although the definition of the auxiliary RVs $(UTV)$ used in the proof makes
them arbitrarily correlated, the bound~\eqref{eq:theo_OB-rate} only depends on
the \emph{marginal} PDs $p(ux_c)$ and $p(tv\vert \hat{y}_s)$. Consequently, we
can restrict the set of possible joint PDs to~\eqref{eq:theo-OB-pmf}, i.e.,
independent source and channel variables, and still achieve the maximum.

The bound on the cardinality of the alphabets $\cU$, $\cT$, and $\cV$ follow from Fenchel--Eggleston--Carath{\'e}odory's theorem and the standard cardinality bounding technique~\cite[Appendix C]{gamal_network_2011}; therefore their proof is omitted.
This concludes the proof of Theorem~\ref{theo-OB}.
\hfill\IEEEQED

\section{Proof of Theorem~\ref{theo-KA-OB} (SK Rate Upper Bound)}
\label{sec:Proof_KA-OB}

Let $R_k$ be an achievable strong secret key rate according to Definition~\ref{def:sk_rate}.
Then, for $\epsilon>0$ and sufficiently large $n$, 
there exist functions $\varphi_i(\cdot)$, $\psi_a(\cdot)$, and $\psi_b(\cdot)$ such that
\begin{subequations}\label{eq:KA-OB_functions}
\begin{align}
X_{ci}    &= \varphi_i(R_r, \hat{Y}_s^{i-1} ) , \label{eq:KA-OB_functions1}\\
K_n       &= \psi_a(R_r, \hat{Y}_s^n) , \\
\hat{K}_n &= \psi_b( Y_s^n, Y_c^n ) ,
\end{align}
\end{subequations}
which verify
\begin{align}
\PR{ \smash{\hat{K}_n} \neq K_n } &\le \epsilon , \label{eq:KA-OB_cond1}\\
\I{K_n}{Z_s^n Z_c^n} &\leq \epsilon , \label{eq:KA-OB_cond2}\\
nR_k -\H{K_n} &\leq \epsilon , \label{eq:KA-OB_cond3}
\end{align}
where we have dropped the conditioning on the codebook $\mathsf{c}_n$ from~\eqref{eq:KA-OB_cond2}, \eqref{eq:KA-OB_cond3}, and all subsequent calculations for clarity.

This proof follows similar steps as the proof presented in Appendix~\ref{sec:Proof-OB}, thus we only point out the differences.
First consider,
\begin{subequations}
\begin{align}
nR_k &\leq \H{K_n} +\epsilon \nonumber\\
 &\leq \IC{K_n}{Y_s^n}{Y_c^n} -\IC{K_n}{Z_s^n}{Y_c^n} +\IC{K_n}{Y_c^n}{Z_s^n} \nonumber\\
&\quad -\IC{K_n}{Z_c^n}{Z_s^n} +n\epsilon_n \label{eq:KA-OB_rate1a}\\
 &\leq n \big[ \IC{V}{Y_s}{T} -\IC{V}{Z_s}{T} +\I{U}{Y_c} -\I{U}{Z_c} \nonumber\\
&\quad +\epsilon_n \big], \label{eq:KA-OB_rate1b}
\end{align}
\end{subequations}
where
\begin{itemize}
 \item \eqref{eq:KA-OB_rate1a} is obtained using similar steps as those in~\eqref{eq:OB_rate1}; and,
 \item \eqref{eq:KA-OB_rate1b} arises from the same procedure as in~\eqref{eq:OB_rate2}, \eqref{eq:OB_rate3}, and~\eqref{eq:OB_rate3bis} but with $K_n$ instead of $\bM_n$.
\end{itemize}
Letting $n\to\infty$, and taking arbitrarily small $\epsilon_n$, we obtain the bound~\eqref{eq:theo_KA-OB-rate}.

In order to obtain~\eqref{eq:theo_KA-OB-cond}, we use the following Markov chain that is a consequence of~\eqref{eq:KA-OB_functions1},
\begin{equation}
 (Y_s^n Z_s^n) \mkv \hat{Y}_s^n \mkv X_c^n \mkv (Y_c^n Z_c^n) . \label{eq:KA-OB_mkv4}
\end{equation}

Due to the data processing inequality, we have
\begin{equation}
\I{\hat{Y}_s^n}{Y_c^n} \leq \I{X_c^n}{Y_c^n} \leq n \I{X_c}{Y_c} , \label{eq:KA-OB_rate4bis}
\end{equation}
where the last inequality is due to the memorylessness property of the channel.
Next consider,%
\begin{subequations}\label{eq:KA-OB_rate4}
\begin{align}
\I{\hat{Y}_s^n}{Y_c^n} &= \I{\hat{Y}_s^n Y_s^n}{Y_c^n} \label{eq:KA-OB_rate4a} \\
 &\geq \IC{\hat{Y}_s^n}{Y_c^n}{Y_s^n} \nonumber\\
 &= \IC{\hat{Y}_s^n}{K_n Y_c^n}{Y_s^n} -\IC{\hat{Y}_s^n}{K_n}{Y_s^n Y_c^n} \nonumber\\
 &\geq \IC{\hat{Y}_s^n}{K_n Y_c^n}{Y_s^n} -n\epsilon_n \label{eq:KA-OB_rate4b} \\
 &\geq n \big[\IC{\hat{Y}_s}{V}{Y_s} -\epsilon_n \big] , \label{eq:KA-OB_rate4g}
\end{align}
\end{subequations}
where
\begin{itemize}
 \item \eqref{eq:KA-OB_rate4a} follows from the Markov chain~\eqref{eq:KA-OB_mkv4};
 \item \eqref{eq:KA-OB_rate4b} stems from $\HC{K_n}{Y_s^n Y_c^n}\leq n\epsilon_n$ due to~\eqref{eq:KA-OB_functions}, \eqref{eq:KA-OB_cond1}, and Fano's inequality, and $\HC{K_n}{Y_s^n Y_c^n \hat{Y}_s^n}\geq 0$ since $K_n$ is a discrete RV; and,
 \item \eqref{eq:KA-OB_rate4g} is obtained using similar steps as those in~\eqref{eq:OB_rate5} with the proper definition for the auxiliary RV $V$.
\end{itemize}
Putting~\eqref{eq:KA-OB_rate4bis} and~\eqref{eq:KA-OB_rate4} together, letting
$n\rightarrow\infty$, and taking an arbitrarily small $\epsilon_n$, we obtain
the bound~\eqref{eq:theo_KA-OB-cond}.

As in the proof of Theorem~\ref{theo-OB}, we can restrict the cardinality of the auxiliary RVs and the set of possible joint PDs to~\eqref{eq:theo-OB-pmf}, i.e., independent source and channel variables, and still achieve the maximum.
%
This concludes the proof of Theorem~\ref{theo-KA-OB}.
\hfill\IEEEQED

\section{Proof of Lemmas~\ref{lm:KG-Bound} and~\ref{lm:KG-Bound2}}
\label{sec:Proof-Lemma-Bound}

The proof of Lemmas~\ref{lm:KG-Bound} and~\ref{lm:KG-Bound2} are similar, and thus we only present the first one in detail.
The specific differences in the proof of Lemma~\ref{lm:KG-Bound2} are shown later in Appendix~\ref{ssec:Proof-Lemma-Bound_2}.

\subsection{Proof of Lemma~\ref{lm:KG-Bound}}


The proof of this lemma follows largely from the proofs of~\cite[Lemma 22.2]{gamal_network_2011} and~\cite[Lemma 4.1]{bloch_physical_2011}. Unlike those proofs, however, we analyze here the behavior of the codeword $\V{T}_j$ rather than the bin index associated to a source sequence.
In the sequel, we remove the block index $j$ to improve clarity in the presentation.

Let us first introduce the random variable $\Upsilon$, such that
\begin{equation*}
\Upsilon \triangleq \ind{(\V{Q}, \V{U}, \V{X}, \V{\hat{Y}}, \V{Z}) \in \typ{QUX\hat{Y}Z}} .
\end{equation*}
Given the random codebook \cC, the randomness in the codeword \V{T} comes from its index $S$. Then, using the binary variable $\Upsilon$, it follows that,
\begin{align}
\HC{\V{T}}{\cC \V{Q} \V{U} \V{Z}} &= \HC{S}{\cC \V{Q} \V{U} \V{Z}} \nonumber\\
 &\leq 1 + \HC{S}{\cC \V{Q} \V{U} \V{Z} \Upsilon} \nonumber\\
 &\leq 1 + \HC{S}{\cC \V{Q} \V{U} \V{Z}, \Upsilon=1} + nS_1 \epsilon' , \label{eq:lemma-bound_1}
\end{align}
where the last inequality is due to $\PR{\Upsilon=0}\leq\epsilon'$.

Now, for a specific codebook $\cC=\mathsf{c}_n$ (which determines the codewords $\V{Q}=\V{q}$ and $\V{U}=\V{u}$) and a sequence $\V{Z}=\V{z}$, let us define the random variable $S_c$ with distribution
\begin{equation*}
 p_{S_c} \triangleq p_{S\vert \cC=\mathsf{c}_n, \V{Q}=\V{q}, \V{U}=\V{u}, \V{Z}=\V{z}, \Upsilon=1} .
\end{equation*}
Therefore,
\begin{equation}
\H{S_c}=\HC{S}{\cC=\mathsf{c}_n, \V{Q}=\V{q}, \V{U}=\V{u}, \V{Z}=\V{z}, \Upsilon=1} . \label{eq:lemma-bound_2}
\end{equation}
Before proceeding, we note that although $S\in[1:2^{nS_1}]$, the index $S_c$ has only a non-zero probability in a smaller subset of indices given the condition on $\V{U}=\V{u}$, $\V{Z}=\V{z}$, and $\Upsilon=1$. In other words, $S_c\in\cS$ where $\cS=[1:2^{nS_1'}]$ and the average value of $S_1'$ is provided in the following lemma.

\vspace{1mm}
\begin{lemma}\label{lem:lem_bnd_S}
Let $\eta_1>0$ and $\varepsilon_1>0$, and let $\chi_1$ be a function of the codebook $\mathsf{c}_n$ and the sequence \V{z} defined as
\begin{equation}
 \chi_1(\mathsf{c}_n, \V{z}) = \ind{ \big| S_1' -\IC{T}{X\hat{Y}}{UZ} -\epsilon_1 \big| \geq \eta_1 } , 
\end{equation}
where $\epsilon_1$ is defined in~\eqref{eq:KG_rates}.
Then, for sufficiently large $n$, $\PR{\chi_1(\cC, \V{Z})=1 \mid \Upsilon=1}\leq \varepsilon_1$.
\end{lemma}
\begin{IEEEproof}
According to the codebook generation procedure from Appendix~\ref{ssec:KG_code}, the expected number of sequences $\V{t}\in\mathsf{c}_n$ such that $\V{t}\in\typ{T\vert \V{q}\V{u}\V{z}}$ is $\bE_{\cC \V{Z}} [|\cS|] = 2^{n(S_1-\alpha)}$
where
\begin{equation*}
 \alpha = -\frac{1}{n}\log\frac{|\typ{T\vert \V{q}\V{u}\V{z}}|}{|\typ{T\vert \V{q}}|} .
\end{equation*}
for some $(\V{q},\V{u},\V{z})\in\typ{QUZ}$.
If we calculate the variance of $|\cS|$, we may then use Chebyshev's inequality to bound the values of $|\cS|$
\begin{equation}
 \PR{ \big| |\cS| -\bE_{\cC \V{Z}} [|\cS|] \big| \geq \epsilon\, \bE_{\cC \V{Z}} [|\cS|] } \leq \epsilon^{-2} 2^{-n(S_1-\alpha)} .
 \label{eq:lemma-bound_3}
\end{equation}
The value of $\alpha$ may be bounded using standard bounds for the cardinality of typical sets.
Finally, taking the logarithm in the argument of the probability of~\eqref{eq:lemma-bound_3} and with an appropriate definition of $\eta_1$ and $\varepsilon_1$, we recover the lemma's statement.
\end{IEEEproof}
\vspace{1mm}

Continuing from~\eqref{eq:lemma-bound_2}, and due to $\V{Q}$ and $\V{U}$ being deterministic given the codebook \cC,
\begin{align}
\MoveEqLeft[2]
\HC{S}{\cC \V{Q} \V{U} \V{Z}, \Upsilon=1} \nonumber\\
 &= \bE_{\cC \V{Z}} \big[ \H{S_c} \big] \nonumber\\
 &\leq \bE_{\cC \V{Z}} \big[ \H{S_c} \mid \chi_1(\cC, \V{Z})=0 \big] + nS_1 \varepsilon_1 , \label{eq:lemma-bound_3bis}
\end{align}
where the last step follows from Lemma~\ref{lem:lem_bnd_S}.
Due to the symmetry of the random codebook generation and encoding procedure, the probability $p_{S_c}$ is independent of the specific value of the index and it only depends on whether the index belongs or not to \cS.
This is addressed in the following lemma.

\vspace{1mm}
\begin{lemma}\label{lem:lem_bnd_ps}
Let $\epsilon>0$ and $\varepsilon_2>0$, and let $\chi_2$ be a function of the codebook $\mathsf{c}_n$ and the sequence \V{z} defined as
\begin{equation}
 \chi_2(\mathsf{c}_n, \V{z}) = \ind{ \big| p_{S_c}(1) -|\cS|^{-1} \big| \geq \epsilon\, |\cS|^{-1} } . 
\end{equation}
Then, $\PR{\chi_2(\cC, \V{Z})=1 \mid \chi_1(\cC, \V{Z})=0, \Upsilon=1 }\leq \varepsilon_2$ for sufficiently large $n$.
\end{lemma}
\begin{IEEEproof}
See Appendix~\ref{ssec:Proof-Lemma-Bound_ps}.
\end{IEEEproof}
\vspace{1mm}

Therefore,
\begin{subequations}\label{eq:lemma-bound_4}
\begin{align}
\MoveEqLeft[1]
\bE_{\cC \V{Z}} \big[ \H{S_c} \mid \chi_1(\cC, \V{Z})=0 \big] \nonumber\\
 &\leq \bE_{\cC \V{Z}} \big[ \H{S_c} \mid \chi_{\cC \V{Z}} \big] + \varepsilon_2 \log|\cS| \label{eq:lemma-bound_4a}\\
 &= \sum_{s \in \cS} \bE_{\cC \V{Z}} [ - p_{S_c}(s) \log p_{S_c}(s) \mid \chi_{\cC \V{Z}} ] + \varepsilon_2 \log|\cS|  \nonumber\\
 &= |\cS| \, \bE_{\cC \V{Z}}[ - p_{S_c}(1) \log p_{S_c}(1) \mid \chi_{\cC \V{Z}} ] + \varepsilon_2 \log|\cS| \nonumber\\
 &\leq (1+\epsilon) \big[ \log |\cS| -\log (1-\epsilon) \big] + \varepsilon_2 \log|\cS| \label{eq:lemma-bound_4b}\\
 &\leq (1+\epsilon+\varepsilon_2)\, n \big[ \IC{T}{X\hat{Y}}{UZ} +\epsilon_1 +\eta_1 \big] \nonumber \\
 &\quad -(1+\epsilon)\log (1-\epsilon) \label{eq:lemma-bound_4c}\\
 &\leq n \big[ \IC{T}{X\hat{Y}}{UZ} +\epsilon_1 +\eta' \big] , \label{eq:lemma-bound_4d}
\end{align}
\end{subequations}
where
\begin{itemize}
 \item \eqref{eq:lemma-bound_4a} is due to 
 Lemma~\ref{lem:lem_bnd_ps}, and $\chi_{\cC \V{Z}}$ is shorthand notation for the condition $\{ \chi_1(\cC, \V{Z})=0, \chi_2(\cC, \V{Z})=0 \}$;
 \item \eqref{eq:lemma-bound_4b} follows from bounding $p_{S_c}(1)$ using Lemma~\ref{lem:lem_bnd_ps};
 \item \eqref{eq:lemma-bound_4c} follows from bounding $|\cS|$ using Lemma~\ref{lem:lem_bnd_S}; and,
 \item \eqref{eq:lemma-bound_4d} holds for some $\eta'>0$.
\end{itemize}
Finally, combining~\eqref{eq:lemma-bound_1}, \eqref{eq:lemma-bound_3bis}, and~\eqref{eq:lemma-bound_4}, we obtain
\begin{equation*}
\HC{\V{T}}{\cC \V{Q} \V{U} \V{Z}} \leq n\, [\IC{T}{X\hat{Y}}{UZ} +\epsilon_1 +\eta''] ,
\end{equation*}
where $\eta''= \eta' +n^{-1} +(\epsilon' +\varepsilon_1) S_1$, which concludes the proof of Lemma~\ref{lm:KG-Bound}.
\hfill\IEEEQED

\subsection{Proof of Lemma~\ref{lm:KG-Bound2}}
\label{ssec:Proof-Lemma-Bound_2}


Let us first introduce a new definition%
\footnote{The sequence \V{Q} is omitted in the sequel given the Markov chain $\V{Q}\mkv (\cC \V{U} \V{T})\mkv (\V{V} \V{X} \V{\hat{Y}} \V{Z})$ that arises due to the codebook generation procedure.} 
for the auxiliary random variable $\Upsilon$,
\begin{equation*}
\Upsilon \triangleq \ind{(\V{U}, \V{T}, \V{X}, \V{\hat{Y}}, \V{Z}) \in \typ{UTX\hat{Y}Z}} .
\end{equation*}
Second, we note again that given the random codebook \cC, the randomness in the codeword \V{V} comes from its index $S$.
Third, for a specific codebook $\cC=\mathsf{c}_n$ (which determines the codewords $\V{U}=\V{u}$ and $\V{T}=\V{t}$) and a sequence $\V{Z}=\V{z}$, let us define the random variable $S_c$ with distribution
\begin{equation}
 p_{S_c} \triangleq p_{S\vert \cC=\mathsf{c}_n, \V{U}=\V{u}, \V{T}=\V{t}, \V{Z}=\V{z}, \Upsilon=1} . \label{eq:lemma-bound_20}
\end{equation}
Fourth, we note that although $S\in[1:2^{nS_2}]$, the index $S_c$ has only a non-zero probability in a smaller subset of indices given the condition on $\V{Z}=\V{z}$ and $\Upsilon=1$. In other words, $S_c\in\cS$ where $\cS=[1:2^{nS_2'}]$ and the average value of $S_2'$ is provided in the following lemma.

\vspace{1mm}
\begin{lemma}\label{lem:lem_bnd_S2}
Let $\eta_1>0$ and $\varepsilon_1>0$, and let $\chi_1$ be a function of the codebook $\mathsf{c}_n$ and the sequence \V{z} defined as
\begin{equation}
 \chi_1(\mathsf{c}_n, \V{z}) = \ind{ \big| S_2' -\IC{V}{X\hat{Y}}{UTZ} -\epsilon_2 \big| \geq \eta_1 } , 
\end{equation}
where $\epsilon_1$ is defined in~\eqref{eq:KG_rates}.
Then, for sufficiently large $n$, $\PR{\chi_1(\cC, \V{Z})=1 \mid \Upsilon=1}\leq \varepsilon_1$.
\end{lemma}
\begin{IEEEproof}
It follows similar steps as those in Lemma~\ref{lem:lem_bnd_S}, and thus it is omitted.
\end{IEEEproof}
\vspace{1mm}

Fifth, due to the symmetry of the random codebook generation and encoding procedure, the probability $p_{S_c}$ is independent of the specific value of the index and it only depends on whether the index belongs or not to \cS.
The statement of Lemma~\ref{lem:lem_bnd_ps} holds although the proof involves characterizing the behavior of the index of \V{V} instead of that of \V{T}. The proof is omitted due to its similarity.

Finally, since we are interested in a lower bound of the index of the codeword \V{V}, \eqref{eq:lemma-bound_1}, \eqref{eq:lemma-bound_3bis}, and~\eqref{eq:lemma-bound_4} may be simplified as%
\begin{subequations}\label{eq:lemma-bound_21}
\begin{align}
\MoveEqLeft[1]
\HC{\V{V}}{\cC \V{U} \V{T} \V{Z}} \nonumber\\
 &= \HC{S}{\cC \V{U} \V{T} \V{Z}} \nonumber\\
 &\geq \HC{S}{\cC \V{U} \V{T} \V{Z}, \Upsilon=1} (1-\epsilon') \label{eq:lemma-bound_21b} \displaybreak[2]\\
 &= \bE_{\cC \V{Z}} \big[ \H{S_c} \big] (1-\epsilon') \label{eq:lemma-bound_21c} \displaybreak[2]\\
 &\geq \bE_{\cC \V{Z}} \big[ \H{S_c} \mid \chi_1(\cC, \V{Z})=0 \big] (1-\epsilon') (1-\varepsilon_1) \label{eq:lemma-bound_21d} \displaybreak[2]\\
 &\geq \bE_{\cC \V{Z}} \big[ \H{S_c} \mid \chi_{\cC \V{Z}} \big] (1-\varepsilon) \label{eq:lemma-bound_21e} \displaybreak[2]\\
 &= |\cS| \, \bE_{\cC \V{Z}}[ - p_{S_c}(1) \log p_{S_c}(1) \mid \chi_{\cC \V{Z}} ] (1-\varepsilon) \nonumber\displaybreak[2]\\
 &\geq (1-\epsilon) \big[ \log |\cS| -\log (1+\epsilon) \big] (1-\varepsilon) \label{eq:lemma-bound_21f}\\
 &\geq (1-\epsilon)\, n \big[ \IC{V}{X\hat{Y}}{UTZ} +\epsilon_2 -\eta_1 \big] (1-\varepsilon) \nonumber \\
 &\quad -(1-\epsilon)\log (1+\epsilon)(1-\varepsilon) \label{eq:lemma-bound_21g}\\
 &\geq n \big[ \IC{V}{X\hat{Y}}{UTZ} +\epsilon_2 -\eta' \big] , \label{eq:lemma-bound_21h}
\end{align}
\end{subequations}
where
\begin{itemize}
 \item \eqref{eq:lemma-bound_21b} follows from $\PR{\Upsilon=1}\geq 1-\epsilon'$;
 \item \eqref{eq:lemma-bound_21c} stems from~\eqref{eq:lemma-bound_20} since $\V{u}$ and $\V{t}$ are fixed given the codebook $\mathsf{c}_n$;
 \item \eqref{eq:lemma-bound_21d} is due to $\PR{ \chi_1(\cC, \V{Z})=0 \mid \Upsilon=1 }\geq 1-\varepsilon_1$ according to Lemma~\ref{lem:lem_bnd_S2};
 \item \eqref{eq:lemma-bound_21e} follows from 
 Lemma~\ref{lem:lem_bnd_ps}, $\chi_{\cC \V{Z}}$ as defined in~\eqref{eq:lemma-bound_4a}, and $(1-\varepsilon)=(1-\epsilon') (1-\varepsilon_1) (1-\varepsilon_2)$;
 \item \eqref{eq:lemma-bound_21f} stems from bounding $p_{S_c}(1)$ using Lemma~\ref{lem:lem_bnd_ps};
 \item \eqref{eq:lemma-bound_21g} stems from bounding $|\cS|$ using Lemma~\ref{lem:lem_bnd_S2}; and,
 \item \eqref{eq:lemma-bound_21h} holds for some $\eta'>0$.
\end{itemize}
This concludes the proof of Lemma~\ref{lm:KG-Bound2}.
\hfill\IEEEQED

\subsection{Proof of Lemma~\ref{lem:lem_bnd_ps}}
\label{ssec:Proof-Lemma-Bound_ps}

According to the encoding procedure detailed in Appendix~\ref{ssec:KG_enc}, the index $S$ is chosen uniformly among all the jointly typical codewords or, if there is no jointly typical codeword, uniformly on the whole codebook. However, due to the conditioning on $\V{U}$, $\V{Z}$, and $\Upsilon=1$, we restrict the indices to the set \cS. We may thus characterize $p_{S_c}(1)$ as
\begin{equation}
 p_{S_c}(1) = \sum_{(\V{x},\V{\hat{y}}) \in \typ{X\hat{Y}}} \frac{p(\V{x},\V{\hat{y}})}{ \PR{ \typ{X\hat{Y}} } }\ \Upsilon_{\V{x},\V{\hat{y}}} ,
 \label{eq:lemma-bound_11}
\end{equation}
where
\begin{equation}
 \Upsilon_{\V{x},\V{\hat{y}}} = \frac{\nu_1}{1+\sum_{i=2}^{|\cS|} \nu_i} +|\cS|^{-1} \prod_{i=1}^{|\cS|} (1 -\nu_i) \label{eq:lemma-bound_12}
\end{equation}
and $\nu_i$ is the event that the codeword $\V{t}(i)$ is jointly typical with the pair $(\V{x},\V{\hat{y}})$, i.e.,
\begin{multline*}
 \nu_i \triangleq \mathds{1} \big\{ \V{t}(i) \in \typ{T\vert \V{u}, \V{x},\V{\hat{y}}} \cond \V{t}(i) \in \typ{T\vert \V{q},\V{u},\V{z}}, \\ (\V{q},\V{u},\V{z}) \in \typ{QUZ\vert \V{x},\V{\hat{y}}} \big\} .
\end{multline*}
The first term in~\eqref{eq:lemma-bound_12} distributes the probability of each pair $(\V{x},\V{\hat{y}}) \in \typ{X\hat{Y}}$ uniformly among all the jointly typical codewords, while the second term in~\eqref{eq:lemma-bound_12} distributes this probability uniformly among all codewords in \cS, given that no one was jointly typical with $(\V{x},\V{\hat{y}})$.
It is not hard to see that the expected value of $\nu_i$ is
\begin{equation*}
 \bE_{\cC \V{Z}}[\nu_i] 
 =  \frac{| \typ{T\vert \V{u},\V{x},\V{\hat{y}}} |}{| \typ{T\vert \V{q},\V{u},\V{z}} |} 
 \triangleq \gamma ,
\end{equation*}
for some $(\V{q},\V{u},\V{x},\V{\hat{y}},\V{z})\in\typ{QUX\hat{Y}Z}$.

The expected value of~\eqref{eq:lemma-bound_11} depends on the behavior of $\Upsilon_{\V{x},\V{\hat{y}}}$. Each $\nu_i$ is a Bernoulli RV with $\bE_{\cC \V{Z}}[\nu_i]=\gamma$ and it is independent of the other $\nu_i$'s. Let us define
\begin{equation*}
 \nu = \sum\nolimits_{i=2}^{|\cS|} \nu_i ,
\end{equation*}
then $\nu$ is a Binomial RV, and thus, for $j\in[0:|\cS|-1]$,
\begin{equation*}
 p_\nu(j) = \binom{|\cS|-1}{j} \gamma^j (1-\gamma)^{|\cS|-1-j} .
\end{equation*}
After some manipulations, it is possible to show that
\begin{equation*}
 \bE_{\cC \V{Z}}\!\left[ \!\frac{1}{1+\nu} \right] = \frac{1-(1-\gamma)^{|\cS|}}{\gamma\, |\cS|} .
\end{equation*}
Hence,
\begin{equation*}
 \bE_{\cC \V{Z}}[\Upsilon_{\V{x},\V{\hat{y}}}]
 = \bE_{\cC \V{Z}}\!\left[ \frac{\nu_1}{1+ \nu} +\frac{1}{|\cS|} \prod_{i=1}^{|\cS|} (1 -\nu_i) \right]
 = \frac{1}{|\cS|} ,
\end{equation*}
and consequently, the expected value of~\eqref{eq:lemma-bound_11} is
\begin{equation*}
 \bE_{\cC \V{Z}}[p_{S_c}(1)]
 = \bE_{\cC \V{Z}}[\Upsilon_{\V{x},\V{\hat{y}}}]
 = |\cS|^{-1} .
\end{equation*}

Noting that $\Upsilon_{\V{x},\V{\hat{y}}}$ and $\Upsilon_{\V{x'},\V{\hat{y}'}}$ are independent variables given different pairs of sequences $(\V{x},\V{\hat{y}})$ and $(\V{x'},\V{\hat{y}'})$, and that $(\Upsilon_{\V{x},\V{\hat{y}}})^2\leq \Upsilon_{\V{x},\V{\hat{y}}}$, we obtain
\begin{equation*}
 \bE_{\cC \V{Z}}[(p_{S_c}(1))^2]
 \leq 2^{-n[\H{X\hat{Y}}-\xi]} |\cS|^{-1} + |\cS|^{-2} ,
\end{equation*}
for some $\xi>0$. Therefore,
\begin{equation*}
 \textnormal{Var}[p_{S_c}(1)] \leq 2^{-n[\H{X\hat{Y}}-\xi]} |\cS|^{-1} ,
\end{equation*}
and in view of Chebyshev's inequality,
\begin{align*}
\MoveEqLeft[2]
\PR{ \big| p_{S_c}(1) -|\cS|^{-1} \big| \geq \epsilon\, |\cS|^{-1} } \nonumber\\
 &\leq \epsilon^{-2} 2^{-n[\H{X\hat{Y}} -\xi]} |\cS| \\
 &\leq \epsilon^{-2} 2^{-n[\H{X\hat{Y}} -\IC{T}{X\hat{Y}}{UZ} -\epsilon_1 -\eta_1 -\xi]} \\
 &= \epsilon^{-2} 2^{-n[\I{UZ}{X\hat{Y}} +\HC{X\hat{Y}}{UTZ} -\epsilon_1 -\eta_1 -\xi]} ,
\end{align*}
where the last inequality follows from Lemma~\ref{lem:lem_bnd_S}.
This concludes the proof of Lemma~\ref{lem:lem_bnd_ps}.
\hfill\IEEEQED

\section{Proof of Lemma~\ref{lm:KG-Fano}}
\label{sec:Proof-Lemma-Fano}

Let us modify the problem definition and then extend the scheme of Theorem~\ref{theo-KG} by introducing a virtual receiver. For each transmission block $j$, this new receiver observes the same channel output $\V{Z}_j$ as the eavesdropper, but it has also perfect access to the codewords $\V{Q}_j$, $\V{U}_j$, and $\V{T}_j$ as well as the indices $L_{2j}$ and $K_j$. In this new setup, we require the virtual receiver to decode the codeword $\V{V}_j$ in each block $j$.

With a slight abuse of notation, we know that according to the codebook generation procedure from Appendix~\ref{ssec:KG_code} and conditioned on the codewords $\V{U}_j$ and $\V{T}_j$, there are $2^{nS_2}$ codewords $\V{V}(L_{2j}, K_j, S_{dj})$. The dummy index $S_{dj}$ represents the position of codeword $\V{V}$ inside the sub-bin $K_j$ and, given the decoding step~\ref{it:dec_v} in Appendix~\ref{ssec:KG_dec}, it is correctly decoded by the legitimate decoder.
Therefore, if we redefine the probability of error for this \emph{enhanced} WTC-GF as
\begin{align*}
\mathsf{P}'_{\!e}(\mathsf{c}_n) \triangleq \PR{ (\hat{\bM}^b, \hat{S}_d^{b}) \neq (\bM^b, S_d^{b}) \textnormal{ or } \hat{S}_d^{b} \neq S_d^{b} \cond \mathsf{c}_n} ,
\end{align*}
we see that a valid code for the enhanced WTC-GF described here is also a valid code for the original WTC-GF.

The extension of Theorem~\ref{theo-KG} is then straightforward; we only need to define the decoding procedure at the virtual receiver. At each block $(j+1) \in [2:b]$, and given $\V{q}_j$, $\V{u}_j$, $\V{t}_j$, $\V{z}_j$, $l_{2j}$, and $k_j$, the virtual receiver looks for the unique index $s_{dj} \equiv \hat{s}$ such that 
\begin{equation*}
 \big( \V{v}(l_{2j},k_j,\hat{s}), \V{q}_j, \V{u}_j, \V{t}_j, \V{z}_j \big) \in \typ{VQUTZ} .
\end{equation*}
Given that $S_2 -\tilde{S}_2 -\bar{S}_2 = \IC{V}{Z}{UT} -\tilde{\epsilon}_2$, the probability of error in decoding is arbitrarily small as $n\to\infty$ if $\delta<\tilde{\epsilon}_2$.
Then, using Fano's inequality, we have
\begin{equation*}
\HC{S_{dj}}{\cC \V{Q}_j \V{U}_j \V{T}_j \V{Z}_j L_{2j} K_j} \leq n\epsilon_n ,
\end{equation*}
where $\epsilon_n$ denotes a sequence such that $\epsilon_n\to 0$ as $n\to\infty$.
The lemma's statement follows from the deterministic relationship between $S_{dj}$ and $\V{V}$ given the codebook, and where we omitted $\V{Q}_j$ due to the Markov chain $\V{Q}_j \mkv (\cC \V{U}_j \V{T}_j \V{Z}_j) \mkv \V{V}$.
This concludes the proof of Lemma~\ref{lm:KG-Fano}.
\hfill\IEEEQED

\section{Proof of Proposition~\ref{prop:erasure_kg}}
\label{sec:Proof-Prop-Erasure}

We proceed to bound from above expressions~\eqref{eq:KG_region1_rate} and~\eqref{eq:KG_region2_rate}, for which we will make use of the variables defined in~\eqref{eq:erasure_rv}. We then show that these upper bounds are achievable with a specific set of random variables.
In particular, an upper bound of~\eqref{eq:KG_region1_rate1} is given by
\begin{multline}
 R_{KG_1} \leq \IC{U}{Y}{Q} -\IC{U}{Z'S}{Q} \\ +\IC{V}{Y}{UT} -\IC{V}{Z'S}{UT} , \label{eq:proof_erasure_1}
\end{multline}
since $\IC{U}{T}{QZ'S}\geq 0$ and $\IC{V}{XS}{UY}$ might be larger than $\I{Q}{Y}$.
Consider now the first two terms on the right-hand side of~\eqref{eq:proof_erasure_1},
where we note that we may add the auxiliary variables $S\triangleq\mathds{1}\{ Y=e \}$ and $S_E\triangleq\mathds{1}\{ Z'=e \}$ alongside $Y$ and $Z'$ without increasing the mutual informations,
\begin{subequations}\label{eq:proof_erasure_2}
\begin{align}
\MoveEqLeft[2]
\IC{U}{Y}{Q} -\IC{U}{Z'S}{Q} \nonumber \\
 &= \IC{U}{YS}{Q} -\IC{U}{Z'SS_E}{Q} \nonumber\\
 &= \IC{U}{Y}{QS} -\IC{U}{Z}{QSS_E} \label{eq:proof_erasure_2a} \\
 &= \IC{U}{X}{Q,S=0}(1-\delta) \nonumber\\
 &\quad -\IC{U}{X}{QS,S_E=0} (1-\delta_E) \nonumber\\
 &= \IC{U}{X}{Q}(\delta_E-\delta) \label{eq:proof_erasure_2b} \\
 &\leq \I{U}{X}(\delta_E-\delta) \label{eq:proof_erasure_2c} ,
\end{align}
\end{subequations}
where
\begin{itemize}
 \item \eqref{eq:proof_erasure_2a} and~\eqref{eq:proof_erasure_2b} are due to $(Q,U,X)$ being independent of $(S,S_E)$; and,
 \item \eqref{eq:proof_erasure_2c} follows from the Markov chain $Q\mkv U\mkv X$ assuming that $\delta_E-\delta\geq 0$. If $\delta_E-\delta< 0$, \eqref{eq:proof_erasure_2} is negative, which means that it is not possible to transmit an unencrypted message, and the rate $R_{KG_2}$ is larger. The reader may later compare the final expressions~\eqref{eq:proof_erasure_5} and~\eqref{eq:proof_erasure_8} to corroborate this claim.
\end{itemize}
Let us concentrate now on the last two terms on the right-hand side of~\eqref{eq:proof_erasure_1},
\begin{subequations}\label{eq:proof_erasure_3}
\begin{align}
\MoveEqLeft[2]
\IC{V}{Y}{UT} -\IC{V}{Z'S}{UT} \nonumber \\
 &= \IC{V}{YS}{UT} -\IC{V}{Z' S S_E}{UT} \nonumber\\
 &= \IC{V}{Y}{UTS} -\IC{V}{Z' S_E}{UTS} \nonumber\\
 &= \IC{V}{Y}{UTS} -\IC{V}{Z'}{UTS S_E} \label{eq:proof_erasure_3a} \displaybreak[2]\\
 &= \IC{V}{X}{UT,S=0}(1-\delta) \nonumber\\
 &\quad -\IC{V}{X}{UTS, S_E=0} (1-\delta_E) \nonumber\\
 &= \IC{V}{X}{UT,S=0}(1-\delta)\delta_E \nonumber\\
 &\quad -\IC{V}{X}{UT,S=1} (1-\delta_E)\delta \label{eq:proof_erasure_3b} \displaybreak[2]\\
 &\leq \IC{V}{X}{UT,S=0}(1-\delta)\delta_E \label{eq:proof_erasure_3c}\\
 &\leq \HC{X}{UT,S=0}(1-\delta)\delta_E \nonumber\\
 &\leq \HC{X}{U}(1-\delta)\delta_E ,
\end{align}
\end{subequations}
where
\begin{itemize}
 \item \eqref{eq:proof_erasure_3a} and~\eqref{eq:proof_erasure_3b} are due to $(U,X,T,V,S)$ being independent of $S_E$; and,
 \item \eqref{eq:proof_erasure_3c} stems from the non-negativity of the mutual information. 
\end{itemize}

On the other hand, an upper bound of~\eqref{eq:KG_region1_rate2} is given by
\begin{equation}
R_{KG_1} \leq \IC{U}{Y}{Q} = \IC{U}{X}{Q} (1-\delta) \leq \I{U}{X} (1-\delta) , 
\label{eq:proof_erasure_4}
\end{equation}
which follows similar steps as~\eqref{eq:proof_erasure_2}.
Therefore, joining~\eqref{eq:proof_erasure_1}--\eqref{eq:proof_erasure_4}, the rate $R_{KG_1}$ may be bounded from above by
\begin{multline}
 R_{KG_1} \leq \max_{p(ux)}\, \min \big\{ \I{U}{X}(\delta_E-\delta) +\HC{X}{U} (1-\delta)\delta_E,\\ \I{U}{X}(1-\delta) \big\} ,
\label{eq:proof_erasure_5}
\end{multline}
which is indeed achievable by selecting the following set of variables:
\begin{align}
  T &= Q = \emptyset &
  & \textnormal{ and } &
  V &= %
\begin{cases}
   X           & \text{if } S=0 \\
   \emptyset   & \text{if } S=1 . 
  \end{cases} \label{eq:proof_erasure_6}
\end{align}
Given that $0\leq \HC{X}{U}\leq \H{X}\leq 1$, we may rewrite the bound~\eqref{eq:proof_erasure_5} using $\HC{X}{U}=\beta$, $\beta\in[0,1]$, 
\begin{multline*}
 R_{KG_1} \leq \max_{\beta\in[0,1]}\, \min \big\{ (1-\beta)(\delta_E-\delta) +\beta (1-\delta)\delta_E,\\ (1-\beta)(1-\delta) \big\} .
\end{multline*}
Upon inspection, we see that the first term increases linearly with $\beta$
while the second one decreases. Therefore, there is a unique maximizer:
\begin{equation}
 R_{KG_1} \leq (1-\delta) \delta_E \frac{\!1 -\delta}{1 -\delta\delta_E} \quad
\textnormal{ for }\quad \beta = \frac{1-\delta_E}{1-\delta \delta_E} .
\label{eq:proof_erasure_7}
\end{equation}

We can proceed similarly for the rate $R_{KG_2}$, by selecting the variables as indicated in~\eqref{eq:proof_erasure_6}, and obtain
\begin{equation}
 R_{KG_2} \leq \max_{p(ux)}\, \min \big\{ \HC{X}{U} (1-\delta)\delta_E,\, \I{U}{X}(1-\delta) \big\} ,
 \label{eq:proof_erasure_8}
\end{equation}
or equivalently:
\begin{equation*}
 R_{KG_2} \leq \max_{\beta\in[0,1]}\, \min \big\{ \beta (1-\delta)\delta_E,\, (1-\beta)(1-\delta) \big\} ,
\end{equation*}
whose maximization gives
\begin{equation}
 R_{KG_2} \leq (1-\delta) \delta_E \frac{1}{1 +\delta_E} \quad \textnormal{ for } \quad \beta = \frac{1}{1 +\delta_E} .
 \label{eq:proof_erasure_9}
\end{equation}
Finally, joining~\eqref{eq:proof_erasure_7} and~\eqref{eq:proof_erasure_9} we obtain the statement of Proposition~\ref{prop:erasure_kg}.
\hfill\IEEEQED

\section*{Acknowledgment}

The authors are grateful to Prof.~Sheng Yang for valuable discussions at the early stage of this work.
The authors would also like to thank the Associate Editor and the anonymous reviewers for their constructive and helpful comments on the earlier version of the paper, which helped to improve the manuscript.

\bibliographystyle{IEEEtran}
\bibliography{IEEEabrv,biblio}

\end{document}